\documentclass{aa}
\usepackage[varg]{txfonts}
\usepackage{graphicx}
\usepackage{subcaption}
\usepackage{amsmath}
\usepackage{hyperref}
\usepackage{enumitem}
\usepackage{soul}
\hypersetup{
    colorlinks=true,
    linkcolor=blue,
    urlcolor=cyan,      
    citecolor=magenta,
    }
\bibpunct{(}{)}{;}{a}{}{,}

\begin{document}

\title{Diversities and similarities exhibited by multi-planetary systems and their architectures}
\subtitle{I. Orbital spacings}
\titlerunning{Multi-planetary systems -- Orbital spacings}
\author{Alexandra Muresan\inst{1}
\and Carina M. Persson\inst{2}
\and Malcolm Fridlund\inst{2,3}
}
\institute{Department of Space, Earth and Environment, Chalmers University of Technology, Chalmersplatsen 4, 412 58 Gothenburg, Sweden.
\and Department of Space, Earth and Environment, Chalmers University of Technology, Onsala Space Observatory, 439 92 Onsala, Sweden.
\and Leiden Observatory, University of Leiden, PO Box 9513, 2300 RA, Leiden, The Netherlands.
}
\date{}

\abstract
{The rich diversity of multi-planetary systems and their architectures is greatly contrasted by the uniformity exhibited within many of these systems. Previous studies have shown that compact \textit{Kepler} systems tend to exhibit a peas-in-a-pod architecture: Planets in the same system tend to have similar sizes and masses and be regularly spaced in orbits with low eccentricities and small mutual inclinations.
This work extends on previous research and examines a larger and more diverse sample comprising all the systems with a minimum of three confirmed planets, resulting in 282 systems and a total of 991 planets. We investigated the system architectures, focusing on the orbital spacings between adjacent planets as well as their relationships with the planets' sizes and masses. We also quantified the similarities of the sizes, masses, and spacings of planets within each system, conducting both intra- and inter-system analyses.
Our results corroborate previous research showing that planets orbiting the same star tend to be regularly spaced and that pairs of adjacent planets with radii $< 1\, R_\oplus$ predominantly have orbital period ratios (PRs) smaller than two. In contrast to other studies, we identified a significant similarity of adjacent orbital spacings not only at \mbox{PRs < 4} but also at \mbox{1.17 < PRs < 2662}.
For the systems with transiting planets, we additionally found that the reported correlation between the orbital PRs and the average sizes of adjacent planets disappears when planet pairs with $R < 1\, R_\oplus$ are excluded. Furthermore, we examined the data for possible correlations between the intra-system dispersions of the orbital spacings and those of the planetary radii and masses. Our findings indicate that these dispersions are uncorrelated for the systems in which all the pairs of adjacent planets have \mbox{PRs < 6}, and even for the compact systems where all \mbox{PRs < 2}. Notably, planets in the same system can be similarly spaced even if they do not have similar masses or sizes.}

\maketitle

\section{Introduction} \label{sect:introduction}
Over the past 30 years, knowledge and achievements in exoplanetary science have increased at an accelerated rate. Technological advancements have empowered scientists to detect, characterise, and simulate an extraordinary plethora of both single- and multiple-planet systems. Surprisingly, many of these systems have astonishing yet unusual properties compared to the Solar System. Exceptionally, the \textit{Kepler} space mission \citep{borucki} has revolutionised astronomy and led to the discovery of 3321 out of all the 5747 confirmed exoplanets to date.\footnote{\url{https://exoplanetarchive.ipac.caltech.edu}. Accessed on 2024-09-06.} This has enabled population-level planetary characterisations \citep[e.g.][]{howard, mulders, otegi20} and statistics \citep{petigura13, zhu21, hsu}. Completely new and, at the same time, the most prevalent types of exoplanets are super-Earths (substantially rocky planets with radii $1.1\, R_\oplus \lesssim R \lesssim 1.8\, R_\oplus$) and sub-Neptunes (mainly gas and ice planets with radii $1.8\, R_\oplus \lesssim R \lesssim 3.5\, R_\oplus$). They are frequently found in compact multi-planetary systems at orbital periods ranging from mere hours to several months \citep{lissauer11, lissauer14, fabrycky14, winn15}. 

The remarkable myriad of planets detected thus far provides us with the opportunity to study the variety of planet and system properties. The architecture of a system describes how the planets are arranged within the system and how their orbital and physical parameters are distributed. In addition, a system's architecture may also provide valuable insights into the multitude of processes that have shaped the system. This may, thereby, enable improvement of the current models of planet formation and evolution.

\begin{figure*}
\centering
\includegraphics[width=17cm]{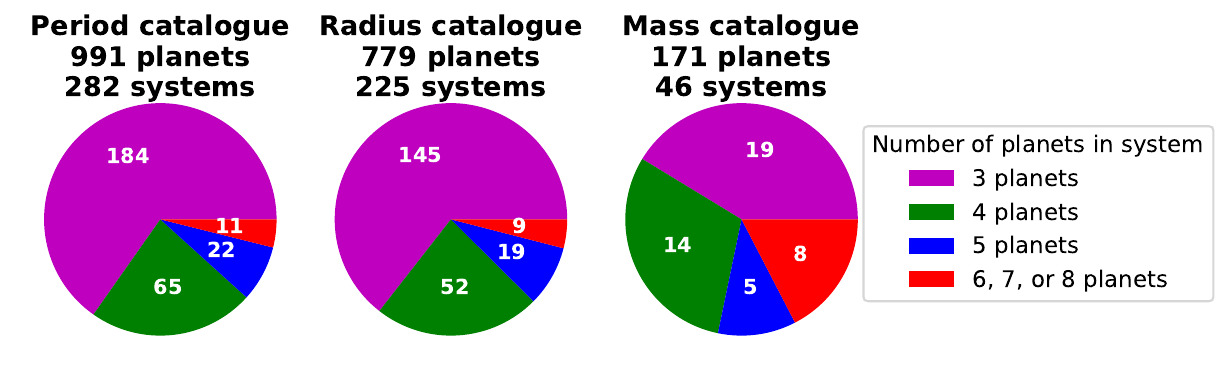}
\caption{Distribution of all the 282, 225, and 46 multi-planetary systems in the period, radius, and mass catalogue, respectively. In white is written the number of systems with any of the four observed planet multiplicities: systems with three (magenta); four (green), five (blue), and either six, seven, or eight (red) planets. Each of these systems has minimum three confirmed planets orbiting a single star, and all its planets have orbital period values. Additionally, each system in the radius catalogue contains at least three adjacent planets with radii measurements. Equivalently, each system in the mass catalogue has minimum three adjacent planets with true mass values.}
\label{fig:piechart}
\end{figure*}

The rich diversity of all the observed exoplanets and planetary systems is greatly contrasted by the uniformity exhibited within many of the multi-planetary systems (i.e. intra-system uniformity). Planets in the same system tend to have equal sizes and masses \citep[e.g.][]{weiss18a, millholland17, weiss23} and be regularly spaced \citep[e.g.][]{weiss18a, jiang20, weiss23} in orbits with low eccentricities \citep{van_eylen15, xie, hadden} and small mutual inclinations \citep{fang, fabrycky14}. Hence, they are often similar to each other, resembling peas in a pod \citep{weiss18a}. 

It was first reported by \citet{lissauer11} that pairs of adjacent planet candidates in \textit{Kepler} systems tend to possess similar radii. \citet{weiss18a} corroborated this finding by examining multi-planetary systems (each harbouring at least two planet candidates) from the California-\textit{Kepler} Survey (CKS) sample (\citealp{johnson}, CKS II; see also \citealp{petigura17}, CKS I). 
They also reported that in systems with a minimum of three planets, the orbital spacing between two adjacent planets, in terms of their period ratio (PR), tends to be similar to the period ratio of the next pair of adjacent planets for \mbox{PRs < 4}. 
Furthermore, they identified a positive correlation between the PRs and the radii of adjacent planets. 
In particular, small-sized planet pairs with radii of $R < 1\, R_\oplus$ were found to possess typical orbital period \mbox{ratios < 2}. The intra-system similarity in planetary radii and orbital spacings has been shown to encompass the planetary masses as well \citep{millholland17, wang}.

This paper is the first in a series focusing on the diversities and similarities exhibited between multi-planetary systems (inter-system) as well as within each system. In the work presented here, we investigate the architectures of observed multi-planetary systems with respect to the orbital spacings (period ratios) of adjacent planets as well as their relationships with the planets' sizes and masses. We explore the similarities and differences of the orbital spacings both on a system level (in each system individually) as well as on a population level. The latter comprises: i) analysing together all the pairs of adjacent planets from all the systems and ii) conducting inter-system comparisons based on the spacing architecture of each system. 

Previous studies on orbital spacings in observed planetary systems have only focused on systems with transiting planets and candidates detected by \textit{Kepler} \citep[e.g.][]{weiss18a, jiang20, weiss23}. In contrast to these, we examine a much larger sample and only include confirmed planets in order to improve the reliability of the data. Our largest data catalogue comprises all systems with at least three planets, thereby encompassing multiple observation techniques: transit photometry, radial velocities (RVs), transit timing variations (TTVs), and direct imaging. This work analyses a much wider range of planetary systems, including orbital periods up to 170\,000~days and period ratios up to 2662.

The key aspects explored in this work are as follows: i) We study whether the spacing similarity trend holds true for our larger and more heterogeneous data sample, for instance, for systems with non-transiting or larger planets. ii) We also investigate which system- or population-level relationships between orbital spacings and planetary sizes or masses emerge in our sample.

We organised this paper as follows: The data catalogues as well as the distributions of orbital periods and planet radii are presented in Sect.~\ref{sect:ctlg}. We examine the orbital period ratios both on a population and system level in Sect.~\ref{sect:spacings}. The relationships between the orbital spacings and the planetary radii and masses are investigated in Sects.~\ref{sect:size} and \ref{sect:mass}, respectively. The orbital separations in units of mutual Hill radii are explored in Sect.~\ref{sect:hill}. We discuss our main results, including possible effects of observational biases, in Sect.~\ref{sect:discussion} and end with conclusions in Sect.~\ref{sect:conclusion}.

\section{Data catalogues} \label{sect:ctlg}

\begin{figure*}
\sidecaption
\includegraphics[width=12cm]{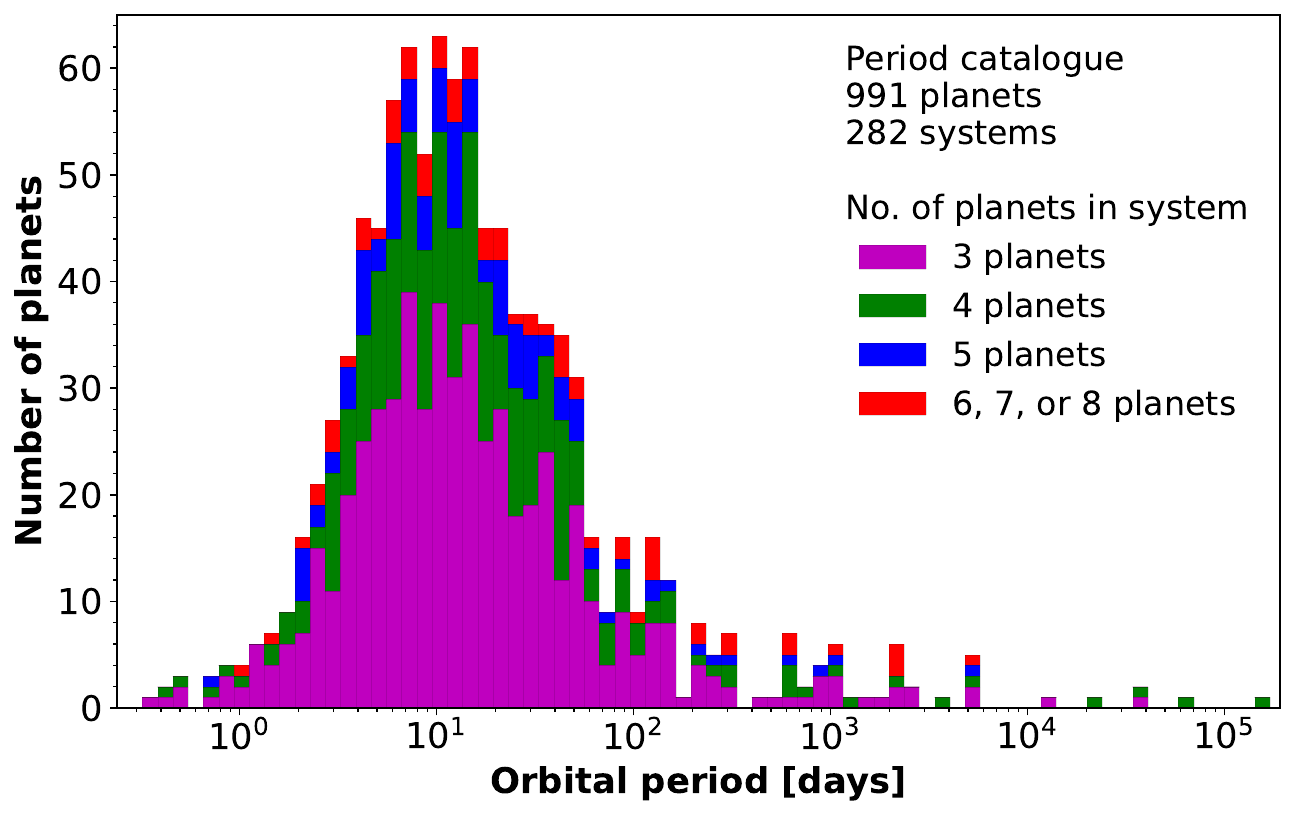}
\caption{Stacked histogram of the orbital periods of all 991 planets in the period catalogue shown for each planet multiplicity: systems with three (magenta), four (green), five (blue), and six to eight (red) detected planets. All four multiplicities have planets spanning the entire range of orbital periods from 0.32 to 170\,000~days. The median of the total distribution is 12.6~days.}
\label{fig:hist_p}
\end{figure*}

\begin{figure}
\resizebox{\hsize}{!}{\includegraphics[width=17cm]{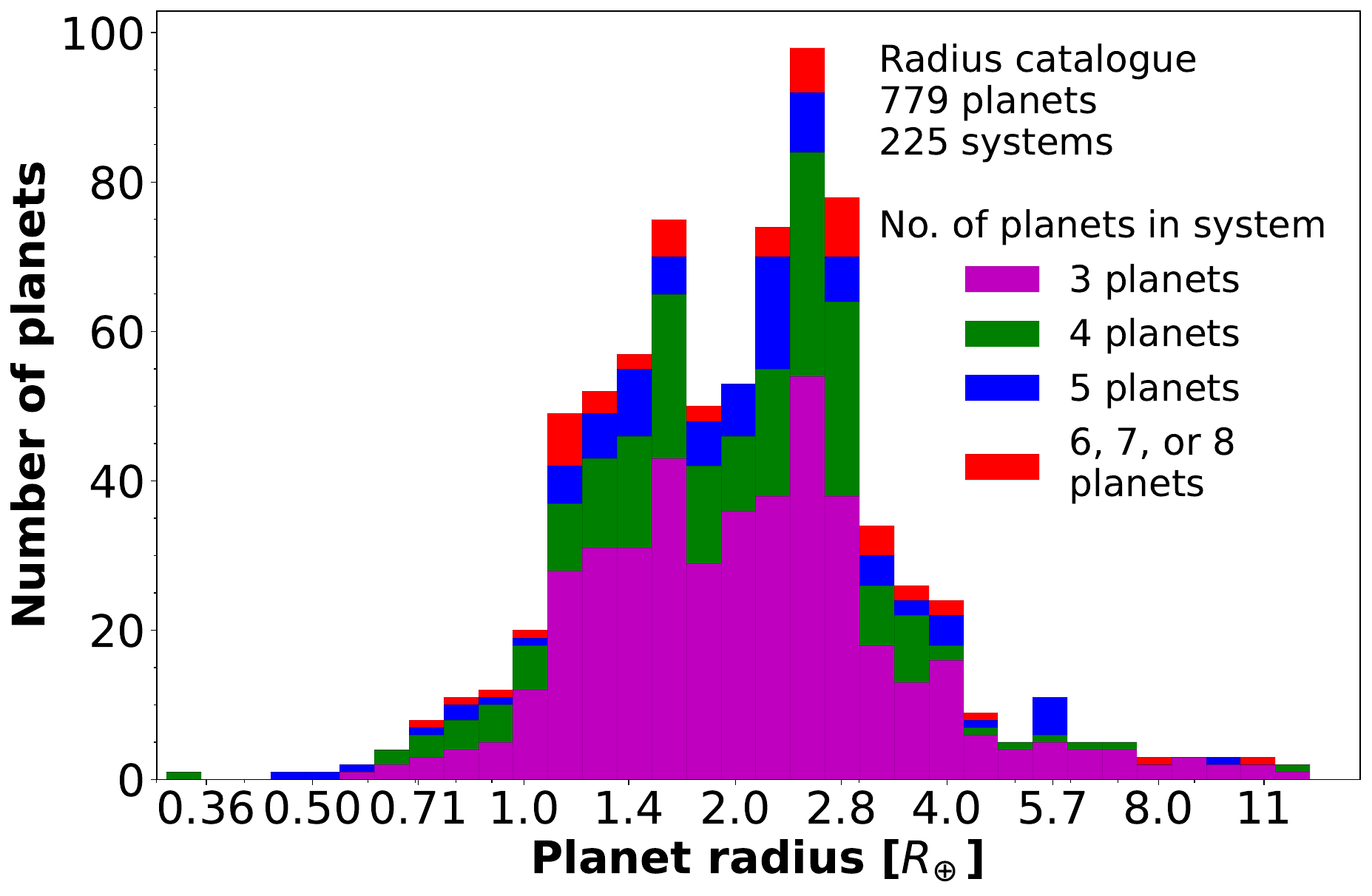}}
\caption{Distribution of the radii of all 779 planets in the radius catalogue shown for each planet multiplicity as a stacked histogram: systems with three (magenta), four (green), five (blue), and six to eight (red) detected planets. The total distribution peaks in the sub-Neptune region at $2.5 - 3.0\, R_\oplus$ while displaying an underabundance at $\approx\!1.8\, R_\oplus$ and an abrupt decline at 3~$R_\oplus$.}
\label{fig:hist_r}
\end{figure} 

\begin{figure}
\resizebox{\hsize}{!}{\includegraphics[width=17cm]{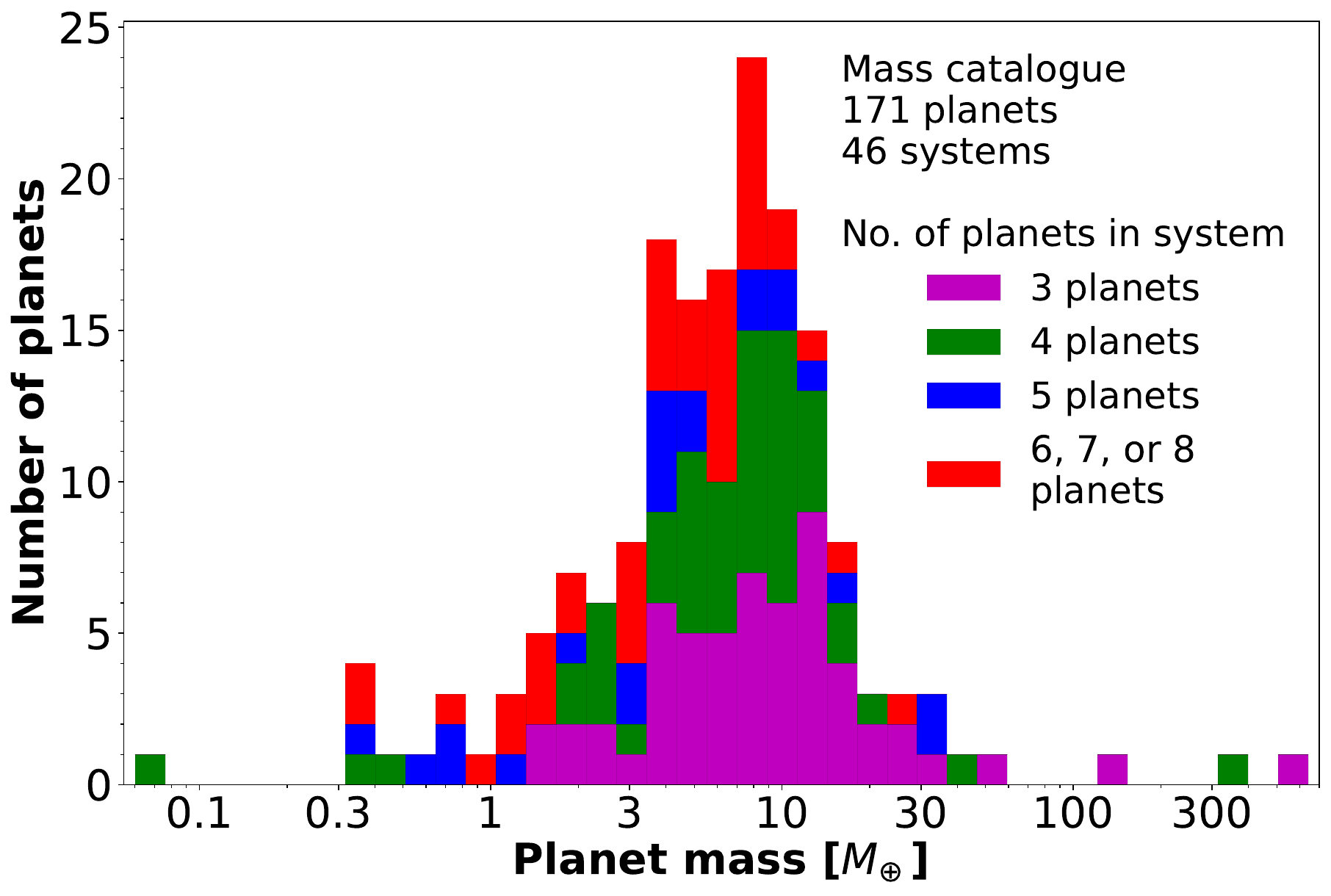}}
\caption{Stacked histogram of the masses of all 171 planets in the mass catalogue shown for each planet multiplicity: systems with three (magenta), four (green), five (blue), and six or seven (red) detected planets. These are the true planetary masses which are obtained either from TTVs or via RVs in conjunction with measurements of the orbital inclinations (Sect.~\ref{sect:ctlg}).}
\label{fig:hist_m}
\end{figure} 

We generated three different catalogues of multi-planetary systems from the planets listed on the NASA Exoplanet Archive.\footnote{See \url{https://exoplanetarchive.ipac.caltech.edu}. Our data catalogues were last updated on 2024-06-28.}
Henceforth, we refer to these samples as the period, radius, and mass catalogues, respectively, as presented in Fig.~\ref{fig:piechart}. We do not include any planet candidates in order to increase the reliability of our samples.
A system is included in the period catalogue if it fulfills the following two criteria:
\begin{enumerate}[label=\alph*), topsep=0.15pt]
    \item It comprises three or more confirmed planets orbiting a single star.
    \item Measurements of the orbital periods of all its confirmed planets are available.
\end{enumerate}
For each system we used the most recent reference that has precise values for the planets' orbital periods as well as their radii and masses when available. The resulting period catalogue is our largest data sample, comprising 282 systems and a total of 991 planets. Only five of these systems host an intermediate planet candidate, which we do not include in our catalogues. Therefore, the measured spacing uniformity in each of these five systems is slightly smaller than it would be if these candidates were taken into account.

A system from this sample is also included for the radius catalogue if it has at least three adjacent planets with available radii values, implying that these planets have been detected via transit photometry. 
As an example, only four of the five planets in the Kepler-82 system are present in the radius catalogue because the outermost planet does not have a radius measurement. This procedure resulted in 225 systems with a total of 779 planets.

Lastly, the mass catalogue contains all the systems from the period catalogue which have minimum three adjacent planets with true planetary mass values. These are derived either via TTVs or from RVs in conjunction with values of the orbital inclinations $i$ from transits. For example, the three-planet systems Kepler-56 and Kepler-88 are not included in the mass catalogue because the outer planet in each of these systems has only $M \sin{i}$ measurements even though the two inner planets have true mass values.
This catalogue contains 46 systems with a total of 171 planets, out of which 170 have radius measurements as well. Kepler-138~e is the only planet in this catalogue which has been discovered by TTVs and has no transit or RV measurements (Fig.~\ref{fig:piechart}). 

The observed planet multiplicity denotes the number of confirmed planets in a system and is either 3, 4, 5, or 6-8. We note for instance that Kepler-82 in the radius catalogue is assigned a planet multiplicity of five even though only its four inner planets are included in this catalogue. 
In total there is only one system with seven planets (TRAPPIST-1), and this is included in all three catalogues. Kepler-90 is the only system with eight planets but is not included in the mass catalogue due to the lack of true mass values for minimum three adjacent planets.

The period catalogue comprises various types of planets and systems, which are useful for studying the diversity of system architectures. The total of 282 systems encompasses planets discovered by different detection techniques: transit photometry (794 planets), RVs (182 planets), TTVs (11 planets), and direct imaging (the HR~8799 system with four planets). This sample is therefore heterogeneous, where each observation method introduces specific biases and limitations. These must be taken into consideration when interpreting the results, as addressed in Sect.~\ref{subsect:bias}. 

On the other hand, the radius catalogue comprises a more homogeneous, although smaller, dataset since it only contains planets discovered by transit photometry. Hence, here we only need to account for the biases and geometrical limitations of this observation method. As seen in Fig.~\ref{fig:piechart}, this catalogue constitutes 225 out of all 282 planetary systems, of which the majority have been detected by the \textit{Kepler} space telescope. Conversely, the mass catalogue contains fewer systems and can only present a very limited view of the diversity of system architectures when exploring the planetary masses in Sects.~\ref{sect:mass} and \ref{sect:hill}.

The distribution of the orbital periods of all the 991 planets in the period catalogue is shown in Fig.~\ref{fig:hist_p} for each of the four planet multiplicities. This sample spans a very wide range of orbital periods. GJ 367~b, with the shortest orbital period of 0.32 days, has been observed with space-based transit photometry, while HR 8799~b, with the longest period of 170\,000~days (465~years) has been detected by direct imaging. The two longest-period planets discovered with the transit method are Kepler-90~h (332 days) and Kepler-150~f (637 days). There are no transiting planets detected at longer periods mainly because the transit probability decreases inversely with the planet's semi-major axis \citep{winn10}. Complementarily, RV surveys have discovered the majority of the long-period planets and only a few of the short-period planets. For example, apart from  the four imaged planets, all 33 planets with orbital periods greater than 637~days have been obtained from RVs. 

The \mbox{0.25-,} \mbox{0.50-,} and \mbox{0.75-}quantiles of the total distribution of orbital periods in the period catalogue are: 5.88, 12.6, and 31.9 days, respectively. Notably, only 10~\% of all planets have \mbox{periods > 100} days.
The median for the three-, four-, five-, and six-to-eight-planet systems is 11.9, 13.3, 12.5, and 16.4~days, respectively. There is no apparent difference between the four multiplicity distributions after normalising them individually. Both the Pearson and Spearman correlation tests indicate that the planet multiplicities are not correlated with the orbital periods. Confirming this conclusion, both the Anderson-Darling and the Kolmogorov-Smirnov tests suggest that all four planet multiplicities can be drawn from the same population, thus corroborating the results of \citet{weiss18b}.

Figure \ref{fig:hist_r} displays a stacked histogram of the radii of all 779 planets in the radius catalogue shown for each planet multiplicity. The radii in this sample range from 0.31~$R_\oplus$ (Kepler-37~b) to 12.64~$R_\oplus$ (WASP-47~b), both of which are found in four-planet systems. The Pearson and Spearman correlation tests indicate that there is no correlation between the planetary radii and the number of planets in a system. Both the Anderson-Darling and the Kolmogorov-Smirnov tests are consistent with this conclusion and indicate no significant difference between the radii distributions of the four planet multiplicities, thereby in agreement with \citet{weiss18b}.

The radius valley is seen around 1.8~$R_\oplus$ and marks the approximate division between super-Earths and sub-Neptunes \citep{fulton, van_eylen18, bean}. The abrupt decline at roughly 3~$R_\oplus$, denoted as the radius cliff, separates sub-Neptunes from Neptunes \citep{kite}. The distribution peaks at $2.5 - 3.0\, R_\oplus$, making sub-Neptunes the most frequent type of planet in our sample. 

The distribution of the masses of all 171 planets in the mass catalogue shown for each planet multiplicity is displayed in Fig.~\ref{fig:hist_m}. The planet masses span a wide range from 0.07~$M_\oplus$ (Kepler-138~b) to 640~$M_\oplus$ (Kepler-30~c).

\begin{figure*}
\sidecaption
\includegraphics[width=12cm]{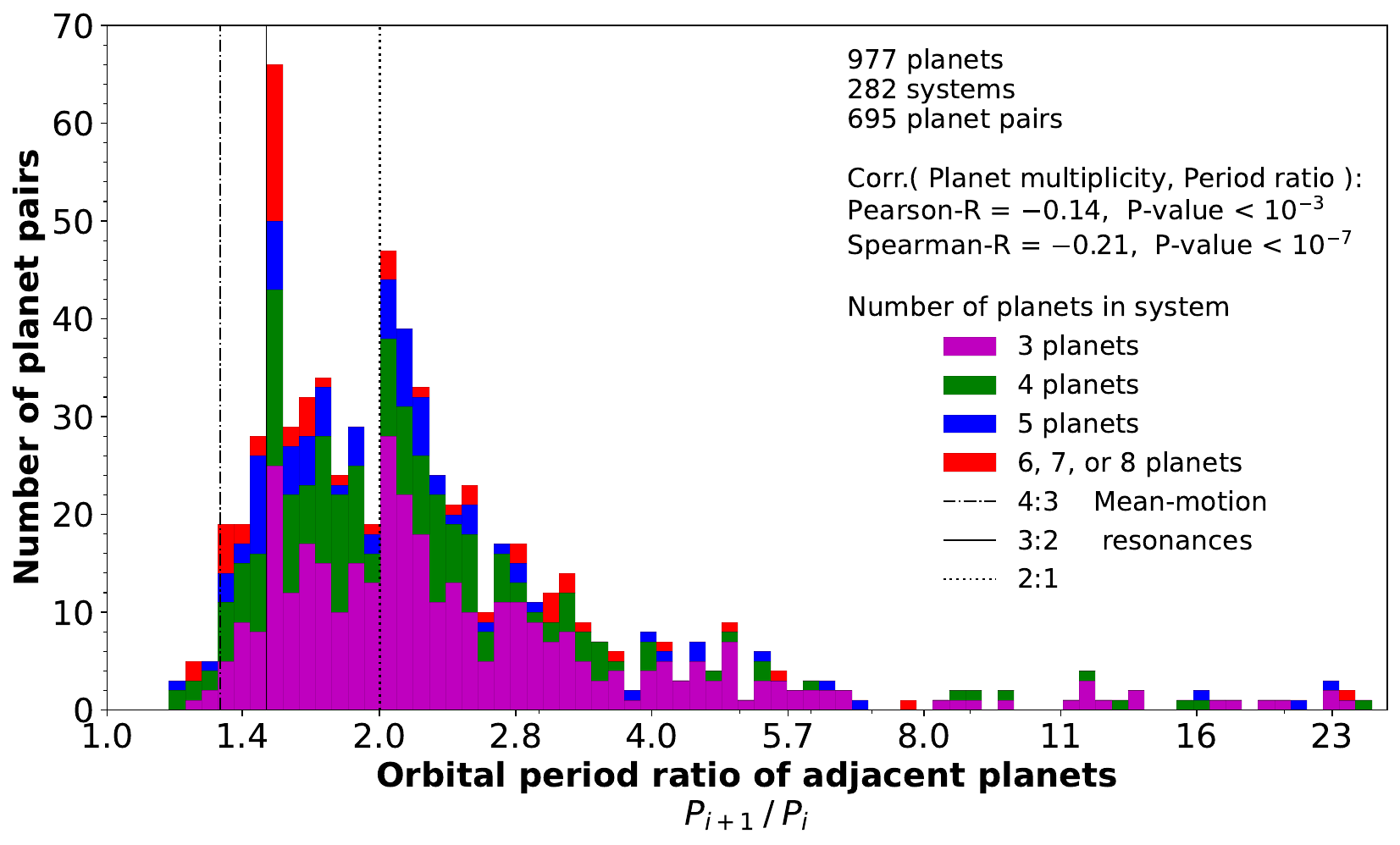}
\caption{Distribution of the orbital PRs of pairs of adjacent planets in the period catalogue up to a PR limit of 25 shown for each planet multiplicity as a stacked histogram: systems with three (magenta), four (green), five (blue), and six to eight (red) detected planets. This sample contains 695 planet pairs, thereby excluding 14 pairs with $25 < \mathrm{PRs} < 2662$. Adjacent planets in higher-multiplicity systems tend to have smaller PRs. There are overabundances of PRs slightly greater than the 4:3, 3:2, and 2:1 MMRs (Sect.~\ref{subsect:diversity}).}
\label{fig:hist_pratio25}
\end{figure*}

\begin{figure*}
\sidecaption
\includegraphics[width=12cm]{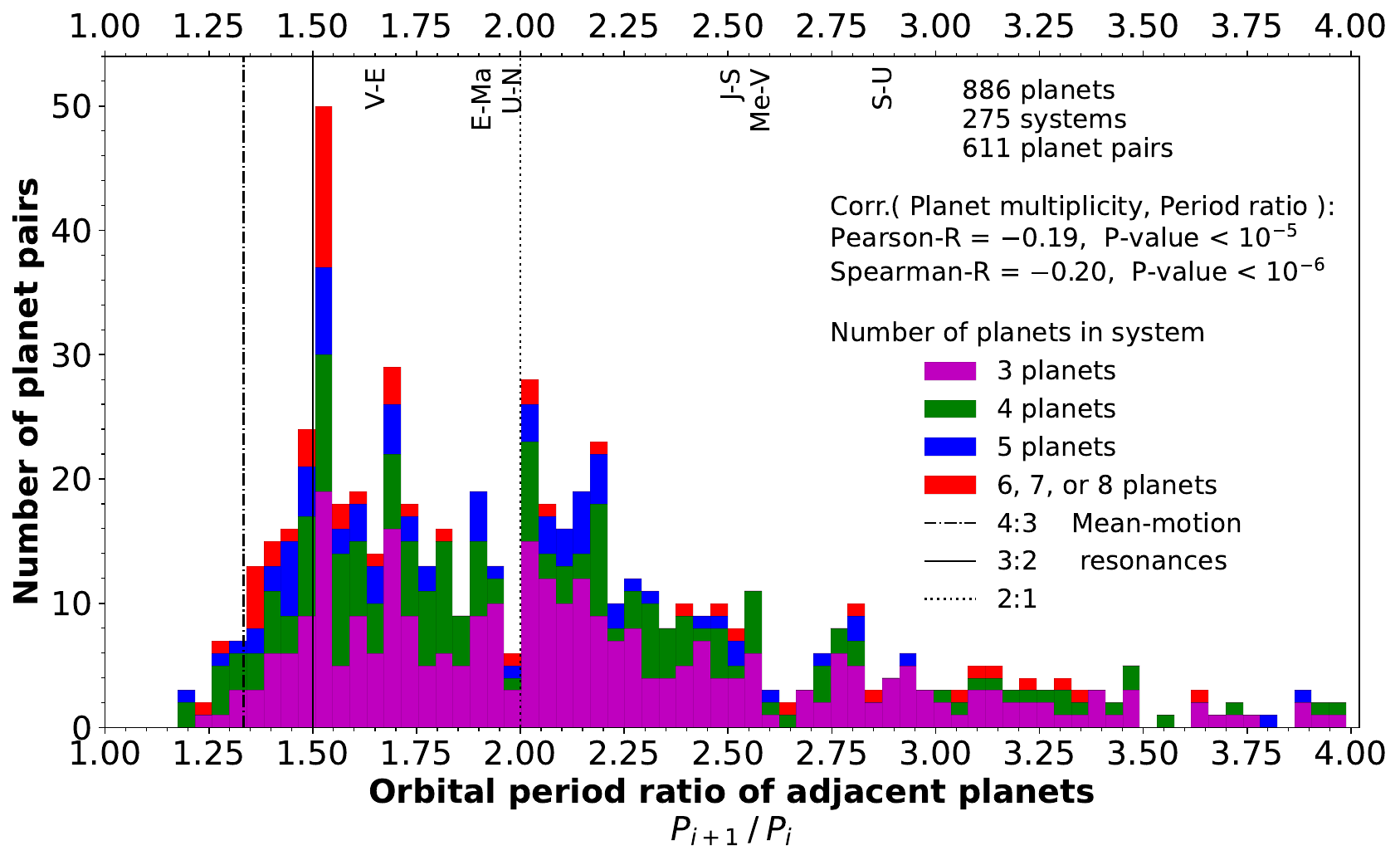}
\caption{Similar to Fig.~\ref{fig:hist_pratio25} but for all 611 pairs of adjacent planets in the period catalogue with $\mathrm{PRs} < 4$ shown on a linear scale. The uppercase letters at the top mark the PR values of all adjacent planet pairs in the Solar System, apart from Mars-Jupiter with $\mathrm{PR} = 6.3$.}
\label{fig:hist_pratio4}
\end{figure*}

\section{Orbital spacings} \label{sect:spacings} 
The relative distance between two adjacent planets is expressed in this paper as the ratio of their orbital periods, $P_{i + 1}/P_i$, where $P_i$ is the orbital period of a given planet in a system, while $P_{i + 1}$ denotes the period of the next adjacent planet, indexed by increasing semi-major axis. We refer to this distance as the orbital spacing or the period ratio (PR). 
We have compared the observed orbital spacings to one another within the same system (on a system level) as well as across many systems (on a population level).

By examining the CKS sample, \citet{weiss18a} reported that the majority of adjacent planet pairs with orbital period \mbox{ratios < 4} possess similar PRs (Sect.~\ref{subsect:similarity}). We note that equal orbital spacings in a system correspond to equal distances in $\log(P)$ since:
\begin{equation} \label{eq:log}
    \log(P_{i + 1}) - \log(P_i) = \log(P_{i + 1}/P_i)\ .
\end{equation}

Due to detection biases and geometrical limitations, long-period planets are prone to remain undiscovered. Hence, this naturally imposes an upper limit on the orbital periods and spacings that we are able to detect. This pertains not only to outer but also inner and intermediate planets since they can be undetectable as well, for instance due to shallow transit depths or low signal-to-noise ratios. Missing an intermediate planet in a system usually creates a large spacing between the two apparently adjacent planets, thus diminishing the intra-system spacing similarity. On the other hand, undetected inner or outer planets might either increase or decrease the intra-system similarity, depending on their spacings relative to their neighbours. Consequently, if one or multiple planets in a system escape detection, the system's true architecture and planet multiplicity are rendered incomplete.

As an example, if a solar-sized star hosts four evenly spaced planets with orbital periods of 2, 20, 200, and 2000~days (corresponding to $\mathrm{PR} = 10$), the geometrical probability of detecting transits of the two outer planets is very low ranging from 0.2~\% to 0.7~\%. Therefore, in a transit survey we might only be able to detect the two inner planets in this system and thereby conclude one of the following three possibilities: (a) This system does not contain a third outer planet due to the large spacing between the two discovered planets; (b) There is an undetected intermediate planet; (c) Due to observational biases, no outer planet is detected, because its orbit has a large semi-major axis or a high inclination.

In order to diminish the effects of these observational biases, we applied different upper limits on the period ratios investigated in this work. This helps decrease the probability of missing outer transiting planets and increase the likelihood of having detected all intermediate planets in a system. In the remainder of this paper, it is explicitly stated when and which PR limits are implemented, particularly limits of 25, four, and two.

\subsection{Diversity of orbital spacings} \label{subsect:diversity}

On a population level, pairs of adjacent planets in observed multi-planetary systems display a wide range of orbital spacings. This is seen in Fig.~\ref{fig:hist_pratio25} which shows a stacked histogram of the distributions of all period \mbox{ratios < 25} of adjacent planets in the period catalogue for each planet multiplicity. 
We applied an upper PR limit of 25 because the region above this threshold is sparsely populated with only 14 planet pairs possessing very high PRs up to 2662. 
A high period ratio may, for instance, be indicative of an undetected intermediate planet or perhaps an asteroid belt similar to the one in the Solar System (Sect.~\ref{sect:discussion}).
The imposed upper limit of \mbox{PR = 25} also serves as a first step towards decreasing the probability of undetected intermediate planets. The resulting sample consists of 282 systems with a total of 977 planets and 695 planet pairs. This is the hitherto largest sample of pairs of confirmed adjacent planets for which the PRs and the PR similarities (Sect.~\ref{subsect:similarity}) have been examined. 

The highest detected orbital period ratio in this selected sample is 24, which is the maximum ratio in both the radius and the mass catalogues as well (the planet pair \mbox{TOI-561 b - c}). As seen in Fig.~\ref{fig:hist_pratio25}, few detected planet pairs have very large orbital spacings. For example, only 84 planet pairs (15~\% of all the pairs in this sample) have $\mathrm{PRs} \geq 4$, and there is an apparent paucity at $\mathrm{PRs} \geq 7$. 
The smallest and highest orbital period ratio in the entire period catalogue without a PR limit is 1.17 and 2662, respectively. The absence of planet pairs with $\mathrm{PRs} < 1.17$ is presumably due to chaotic orbital motion and Lagrange instability caused by overlap of first-order mean motion resonances MMRs \citep{deck}. For low-mass planets with planet-to-star mass ratios of $M_p/M_\star \lesssim 10^{-5}$, this occurs at $\mathrm{PRs} < 1.2$, even if their orbits lie in a region of Hill stability \citep[][Fig. 12]{deck}.

Furthermore, there is a weak negative correlation between the planet multiplicities and the orbital period ratios, meaning that planets tend to be more tightly packed in higher-multiplicity systems. This trend is noticeable in Fig.~\ref{fig:hist_pratio25}, especially for the systems with at least five planets. 
The \mbox{0.25-,} \mbox{0.50-,} and \mbox{0.75-}quantiles of the PR distribution of this sample are: 1.63, 2.06, and 2.76, respectively. 
The median PR for the three-, four-, five-, and six to eight-planet systems is 2.16, 1.93, 1.93, and 1.63, respectively. This corroborates previous findings that planet pairs in high-multiplicity systems possess smaller orbital period ratios \citep{steffen}. Nevertheless, it is important to note that this negative correlation is strengthened by observational biases and limitations, as discussed in Sect.~\ref{subsect:bias}. 

The three strongest first-order mean-motion resonances of 4:3, 3:2, and 2:1 are marked in Fig.~\ref{fig:hist_pratio25}, showing that there are overabundances of planet pairs with ratios slightly greater than these MMRs as well as deficits at PRs slightly less than these MMRs. The most prominent peak in the overall distribution is near the 3:2 MMR, while the greatest through is slightly interior to the 2:1 MMR. These near mean-motion-resonance features have been reported in previous studies of \textit{Kepler} systems and found to be statistically significant \citep[e.g.][]{steffen, lissauer11, fabrycky14}.

To further decrease the probability of undetected planets in systems and to facilitate comparisons with \citet{weiss18a} and \citet{weiss23}, we selected from the period catalogue only the pairs of adjacent planets with $\mathrm{PRs} < 4$. This resulted in 611 planet pairs and is, thus, still the largest sample of PRs of confirmed planets reported in the literature. The PR distribution for each planet multiplicity is displayed in Fig.~\ref{fig:hist_pratio4} on a linear scale. This enabled us to mark the corresponding values for all seven adjacent planet pairs in the Solar System, except for Mars-Jupiter which have a high PR of 6.3. The remaining six pairs have PRs ranging from 1.6 for Venus-Earth to 2.8 for Saturn-Uranus and are, therefore, typical among the exoplanet pairs, as also noted for instance by \citet{malhotra}. By applying an upper PR limit of 4, the planet multiplicity and orbital period ratios are slightly more strongly correlated, leading to Pearson-R~$= -0.19$ as opposed to Pearson-R~$= -0.14$ for the previous sample with $\mathrm{PRs} < 25$.

Furthermore, we computed several sets of both Anderson-Darling and Kolmogorov-Smirnov tests in order to assess whether the four multiplicity distributions of PRs are significantly different or can be drawn from the same population. The following conclusions are the same both when applying no upper PR limit and with a limit of either 4 or 25. The sets of both types of tests indicate that the three-planet systems are drawn from a different distribution of PRs (i.e. skewed towards larger values) than the systems with a minimum of four planets. However, there is not yet sufficient data to study whether this pattern extends to higher multiplicities.

\subsection{Similarity of adjacent orbital spacings} \label{subsect:similarity}

In contrast to the diversity of period ratios exhibited by the pairs of adjacent planets on a population level in Fig.~\ref{fig:hist_pratio25}, the orbital spacings within an individual system tend to be equal \citep[e.g.][]{weiss18a, jiang20}. This means that planets orbiting the same star tend to be equally spaced in log-period (Eq.~\ref{eq:log}). This intra-system uniformity is a key feature of the peas-in-a-pod architecture in addition to the similarities of the planetary sizes and masses as well as the low orbital eccentricities and mutual inclinations in a system. Previous studies analysing \textit{Kepler} data have determined that the spacing similarity trend is not caused by detection biases or limitations and is, therefore, not an artefact of \textit{Kepler}'s discovery efficiency \citep{weiss20, mishra21, weiss18a, he19, jiang20, weiss23, lammers}. Furthermore, population synthesis models of planetary systems have revealed that the intra-system spacing similarity is a typical outcome of system formation and evolution \citep{mishra21, mishra23b, weiss23}. It has also been shown that this spacing trend is diminished in observations due to the biases and limitations of the transit method as well as the completeness of the \textit{Kepler} survey \citep{mishra21}.

In order to investigate the spacing similarity in a system with $N$ planets, the period ratios of all pairs of adjacent planets need to be compared: $P_{i + 1}/P_i$ for $i = 1, 2, ..., N$.
The standard approach employed in previous work is to plot $P_{i + 2}/P_{i + 1}$ against $P_{i + 1}/P_i$ for all adjacent planet pairs and test for a correlation. The points representing a system in which all adjacent planets have equal PRs lie perfectly on top of each other at a certain value on the one-to-one line in such a plot. 

Despite their simplicity, correlation tests are only suitable for large datasets and cannot be applied on a single planetary system (i.e. on a system level).
This is due to the very low number of data points, since $N$ planets result in $N - 1$ planet pairs and only $N - 2$ points to plot. Therefore, in order to exploit the advantages of correlation tests, all the adjacent planet pairs in all the systems should be plotted on the same graph (i.e. on a population level).
The disadvantage, however, is that information about an individual system is lost since we do not know which data points correspond to the same system. In order to measure the similarity on a system level, we therefore employed another method as explained in Sect.~\ref{subsect:intra-system similarity}.

\begin{figure*}
\sidecaption
\includegraphics[width=12cm]{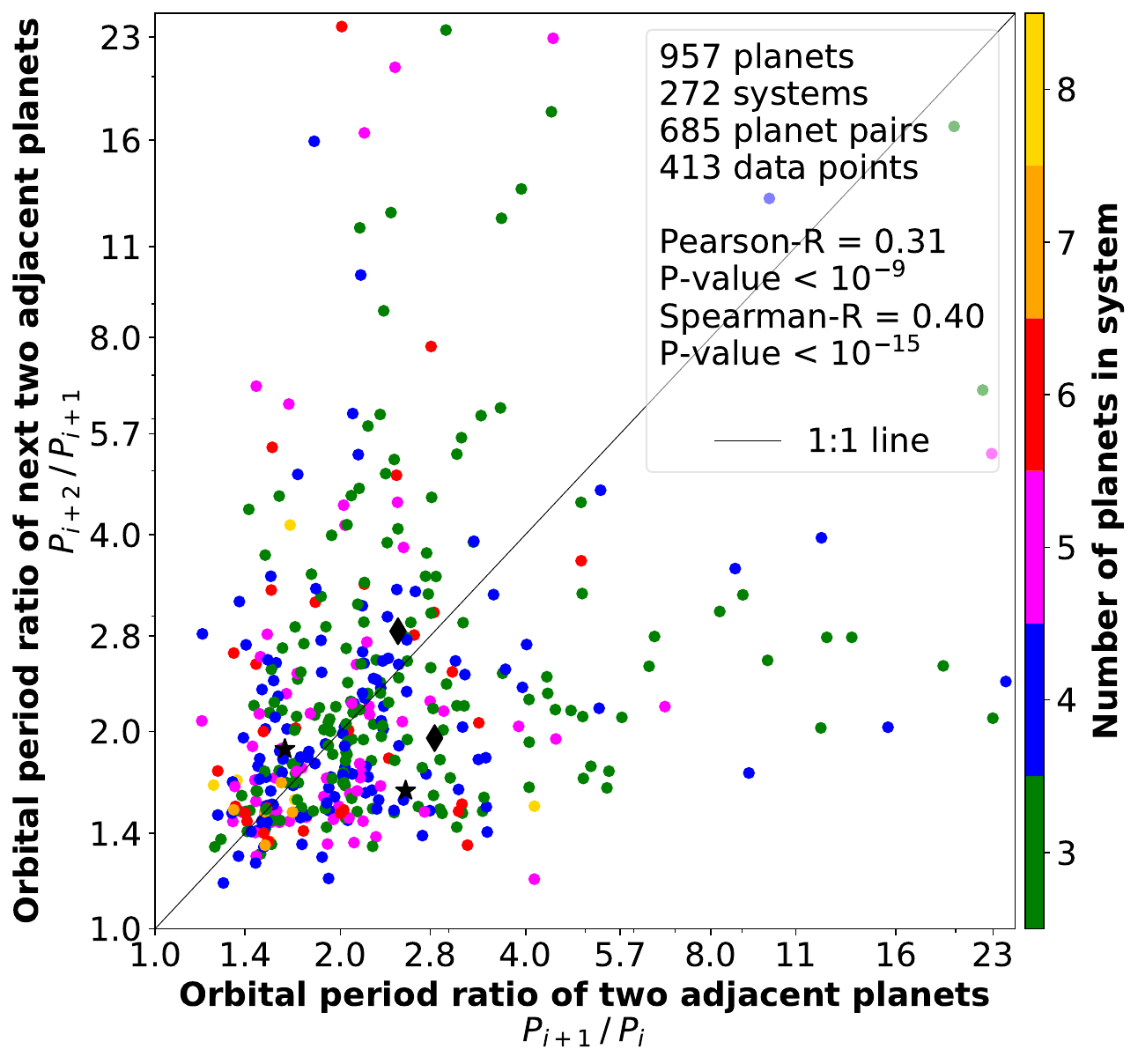}
\caption{Period ratio of a pair of adjacent planets versus the PR of the previous such pair for all 685 pairs in the period catalogue where every two adjacent pairs have \mbox{PRs < 25}. Hence, each data point represents two adjacent planet pairs, resulting in 413 points colour-coded based on the planet multiplicity in each system. There is a positive correlation, although the points are greatly dispersed around the 1:1 line. Points representing planet pairs in a system where there is a perfect spacing similarity lie precisely on top of each other on this line. The two black stars and diamonds mark the planet pairs in the inner and outer Solar System, respectively (Sect.~\ref{subsect:similarity}).}
\label{fig:pratio25}
\end{figure*}

Figure \ref{fig:pratio25} displays the orbital period ratio of a pair of adjacent planets $P_{i + 2}/P_{i + 1}$ vs. the PR of the previous such pair $P_{i + 1}/P_i$ in the period catalogue. We selected the systems with at least two adjacent planet pairs with $\mathrm{PRs} < 25$. This sample is slightly smaller than the one displayed in Fig.~\ref{fig:hist_pratio25} but is still the largest for which the spacing similarity has been examined thus far. It contains 272 systems with a total of 957 planets, 685 planet pairs, and 413 data points. The previously largest sample was analysed by \citet{weiss18a} who selected from the CKS multis the systems with a minimum of three candidates or confirmed planets in which at least two adjacent planet pairs had $\mathrm{PRs} < 4$. This resulted in 104 systems with a total of 373 planets and 165 data points. Their sample is, therefore, significantly smaller than the one we investigated in this work. 

\begin{figure*}
\centering
\begin{subfigure}{.5\textwidth}
  \centering
  \includegraphics[width=7.9cm]{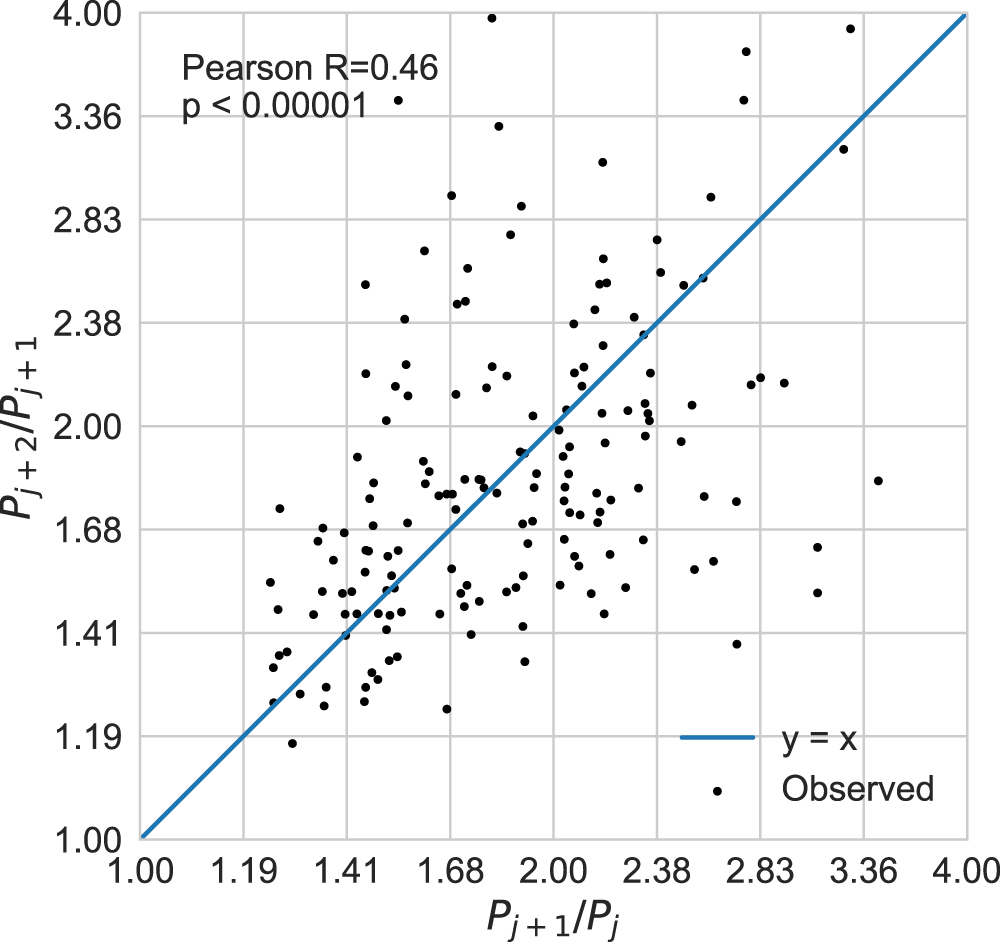}
\end{subfigure}%
\begin{subfigure}{.5\textwidth}
  \centering
  \includegraphics[width=9.1cm]{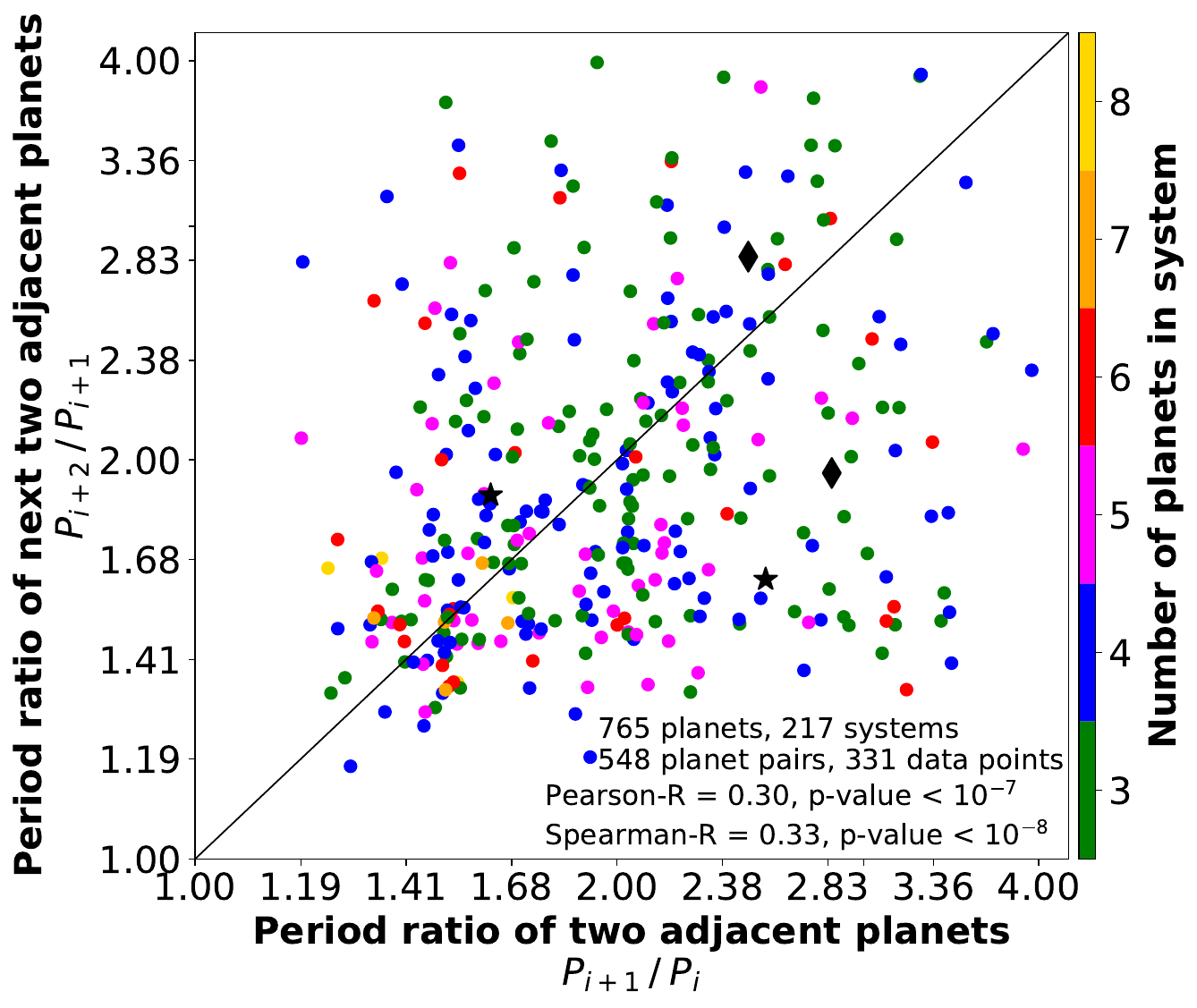}
\end{subfigure}
\caption{As in Fig.~\ref{fig:pratio25} but now only for \mbox{PRs < 4}. Left figure: Graph from \citet{weiss18a} displaying 165 data points from a total of 104 systems and 373 planets. This has been the largest observed sample for which a strong similarity of adjacent spacings has been found. Right figure: Sample from the period catalogue in this work comprising 331 data points from a total of 217 systems, 548 planet pairs, and 765 planets. Although this sample is smaller than that in Fig.~\ref{fig:pratio25}, it is twice as large as that of \citet{weiss18a} and shows a weaker, but more significant, correlation.}
\label{fig:pratio4}
\end{figure*}

Figure \ref{fig:pratio25} also displays the corresponding values for the inner Solar System (the four terrestrial planets) and the outer Solar System (the four giant planets). These planet pairs do not have equal orbital spacings but are, nonetheless, typical among the other pairs, tending to follow the one-to-one line. 

There is a significant positive correlation between the orbital period ratio of a pair of adjacent planets and the PR of the next such pair in the same system. However, the upper limit on the period ratios affects the strength of this correlation. With no upper PR limit for the CKS multis, \citet{weiss20} found a weak positive correlation that was barely significant (\textit{p}-value = 0.003). Notwithstanding this result, \citet{jiang20} identified no correlation at all for the \textit{Kepler} systems with a minimum of three candidates (including confirmed planets) after applying a limit on the planetary radii of $6\, R_\oplus$ and no limit on the period ratios. Recently, also \citet{mamonova} found no correlation between the orbital PRs of adjacent planets in their sample of observed multi-planetary systems. Contrary to the aforementioned studies, our analysis suggests that a correlation does in fact exist even when large orbital period ratios up to 2662 are included. 

Additionally, we investigated how the upper PR limit alters the correlation statistics for our sample. The linear correlation coefficient for the entire period catalogue with $1.17 < \mathrm{PRs} < 2662$ is Pearson-\mbox{R = 0.20}, after which it increases to its maximum value of Pearson-\mbox{R = 0.31} for $1.17 < \mathrm{PRs} < 25$. By further decreasing the upper PR limit, the correlation coefficient fluctuates between 0.26 and 0.31 until $\mathrm{PRs} < 1.7$ at which the \textit{p}-value becomes insignificant. 

It is visible in Fig.~\ref{fig:pratio25} that the majority of points get farther away from the 1:1 line as the period ratios increase, implying that the correlation is diminished at high orbital spacings. Noticeably, there is an absence of points in the upper right corner of this figure. This is primarily caused by observational biases because two adjacent planet pairs are most likely undetectable if they both have very high PRs. In this region, the correlation statistics are, therefore, misrepresented, which is the reason why proper PR limits are needed. 
As seen in Fig.~\ref{fig:pratio25}, especially data points at \mbox{PRs > 4} are largely dispersed relative to the 1:1 line, and a very large orbital spacing in one planet pair is mainly associated with a much smaller spacing in the adjacent pair, since otherwise both pairs might not have been detected.

Therefore, in order to mitigate these effects of observational biases and, thereby, decrease the probability of intermediate and outer undetected planets, we imposed an upper PR limit of 4, as implemented in \citet{weiss18a}. The resulting sample acquired from the period catalogue contains 217 systems with a total of 765 planets and 548 planet pairs, fulfilling that \mbox{PRs < 4} for every two adjacent planet pairs.
Figure~\ref{fig:pratio4} shows the similarity of the adjacent orbital spacings in this sample (331 data points) along with the corresponding plot (165 data points) presented in \citet{weiss18a}. Although the correlation in our sample is weaker than in \citet{weiss18a}, it is more significant and indicates that the peas-in-a-pod spacing pattern at \mbox{PRs < 4} emerges even in a sample twice as large.
 
Another distinctive feature in both Figs.~\ref{fig:pratio25} and \ref{fig:pratio4} is the near symmetry along the one-to-one line. The data points $(x,\,y)$ are approximate reflections of each other across this line. Moreover, this property is independent of the number of planets in the systems. There is no correlation between the planet multiplicity and whether $x < y$, or $y < x$; that is, whether the inner or outer planet pair has a higher period ratio. This near symmetry about the 1:1 line emerges both owing to the peas-in-a-pod planets that lie along this line as well as due to observational biases since a large PR in a planet pair necessitates a much smaller PR in the adjacent pair in order for both pairs to be detected.

Since planets in higher-multiplicity systems tend to be more tightly packed (Sect.~\ref{subsect:diversity}), we focused next exclusively on the systems in the period catalogue with a minimum of four planets. We conducted the same analysis and examined the spacing similarities as in Fig.~\ref{fig:pratio25}. The results indicate that irrespective of the upper PR limit, the correlation between the orbital period ratios of adjacent planet pairs is weaker in this sample that excludes the three-planet systems.

In order to explore how the planets detected only with the RV method influence the spacing similarity trend, we examined the planets in the radius catalogue since they all have been detected by transit photometry. \mbox{TOI-561 b - c} is the planet pair with the highest PR of 24, which is the same as in Fig.~\ref{fig:pratio25} for the period catalogue limited to $\mathrm{PRs} < 25$. The correlation between adjacent PRs has now the coefficient Pearson-R = 0.21 and is, therefore, weaker compared to the previous Pearson-R = 0.31. For the planet pairs with $\mathrm{PRs} < 4$, the Pearson-R is 0.27 and 0.30 for the radius and period catalogue, respectively. These two results have two main implications: First, the planets detected with the RV technique increase the strength of the spacing similarity trend (for $\mathrm{PRs} < 25$). Second, they predominantly affect the correlation at $\mathrm{PRs} > 4$ since the majority of them have larger orbital spacings than the transiting planets.

\subsection{Intra-system similarity of orbital spacings} \label{subsect:intra-system similarity}

While statistical correlation tests, such as the Pearson and Spearman tests, are applicable on large datasets in population-level research, they are not suitable for measuring distinctive trends within an individual system. Therefore, other appropriate tests or measures must be devised in order to examine the intra-system properties and correlations. There are only a few system-level studies reported in the literature \citep{kipping, gilbert, goyal, mishra23a, weiss23, mamonova}. 

In this work we employed a straightforward metric that quantifies the similarity of a specified parameter, as for instance the orbital spacings or the planet masses within an individual system. This facilitates both intra- and inter-system analyses of multi-planetary systems. 
Similar to \citet{weiss23}'s approach, we implemented the fractional dispersion, which is the standard deviation $\sigma_q$ of a quantity $q$ defined as follows:
\begin{equation} \label{eq:sigma}
\begin{split}
    \mathrm{Intra\text{-}system\, dispersion\, of\,} q & = \sqrt{\frac{1}{N} \sum _{i = 1}^{N} {\Bigl\{ \log(q_i) - \log(\overline{q}) \Bigl\}}^2} \\ & = \sqrt{\frac{1}{N} \sum _{i = 1}^{N} {\Biggl\{ \log \left(\frac{q_i}{\overline{q}}\right) \Biggl\}}^2} \ ,
\end{split}
\end{equation}
Where $N$ designates the number of planets in the system and is $\geq 3$. The log-base is set to 10, and $\overline{q}$ denotes the mean of the quantity $q$ in the system.

We applied Eq.~\ref{eq:sigma} on the orbital period ratios ($\sigma_\mathrm{PR}$) as well as the planetary radii ($\sigma_R$; Sect.~\ref{sect:size}), masses ($\sigma_M$; Sect.~\ref{sect:mass}), and Hill radii separations ($\sigma_\Delta$; Sect.~\ref{sect:hill}). $\sigma_q$ defined in this work is dimensionless and can range from 0 up to very large numbers.
Henceforth, $\sigma_\mathrm{PR}$ is referred to as the spacing dispersion or, equivalently, as the dispersion of period ratios or of spacings. If all the planets in the same system resemble peas in a pod, they display a small dispersion and, thereby, a strong intra-system similarity. 

We also quantified the intra-system similarity by computing the coefficient of variation, $C_\mathrm{V} = \sigma(q)/\overline{q}$, which is defined as the standard deviation of a quantity \textit{q} divided by the mean \textit{q} of a dataset \citep{brown}. 
The results from the two metrics utilised in our analyses are consistent with one another, and henceforth we only report on the spacing dispersion $\sigma_\mathrm{PR}$. 

It is important to note that a compact system has planets with short orbital periods but does not implicitly exhibit a low spacing dispersion.
This also means that if adjacent planets in a non-compact system have high but nearly equal period ratios, the resulting spacing dispersion is low. However, due to observational limitations, the outer planets in non-compact systems are prone to remain undetected. 

We computed the spacing dispersions in all 282 systems in the period catalogue in order to identify common architecture trends. The five systems with the lowest dispersions, ranging from 0.00047 to 0.00084, are in increasing order: Kepler-398, Kepler-229, Kepler-271, Kepler-207, and Kepler-289. In the opposite extremity, the five largest dispersions range from 0.85 to 1.99 and belong to the following systems in increasing order: GJ 433, Kepler-88, HD 181433, HD 27894, and HD 153557. It is noteworthy that each of these five systems contain at least one long-period planet detected by RVs (Sect.~\ref{subsect:undetected}). As a comparison, the dispersion of orbital spacings in the inner, outer, and entire Solar System are 0.082, 0.068, and 0.18, respectively. Hence, our host system has an intermediate value and is typical among the discovered multi-planetary systems in terms of the intra-system spacing similarities. However, when comparing only the inner regions of systems with inner transiting planets and minimum one outer giant, the inner Solar System possesses the smallest spacing dispersion (Sect.~\ref{sect:size}).

Disregarding observational biases, from the above examinations of spacing dispersions we can highlight two main results that apply on the observed systems in our catalogue. 
First, the majority of systems with very large dispersions contain one pair of adjacent planets that has a much higher period ratio than all the other pairs in its respective system. Second, there is neither a Pearson nor a Spearman correlation between the spacing dispersions and the planet multiplicity in the systems. This is also in agreement with our results from the Anderson-Darling and Kolmogorov-Smirnov tests, indicating that the dispersion distributions for all four planet multiplicities can be drawn from the same population. 

\begin{figure}
\resizebox{\hsize}{!}{\includegraphics[width=17cm]{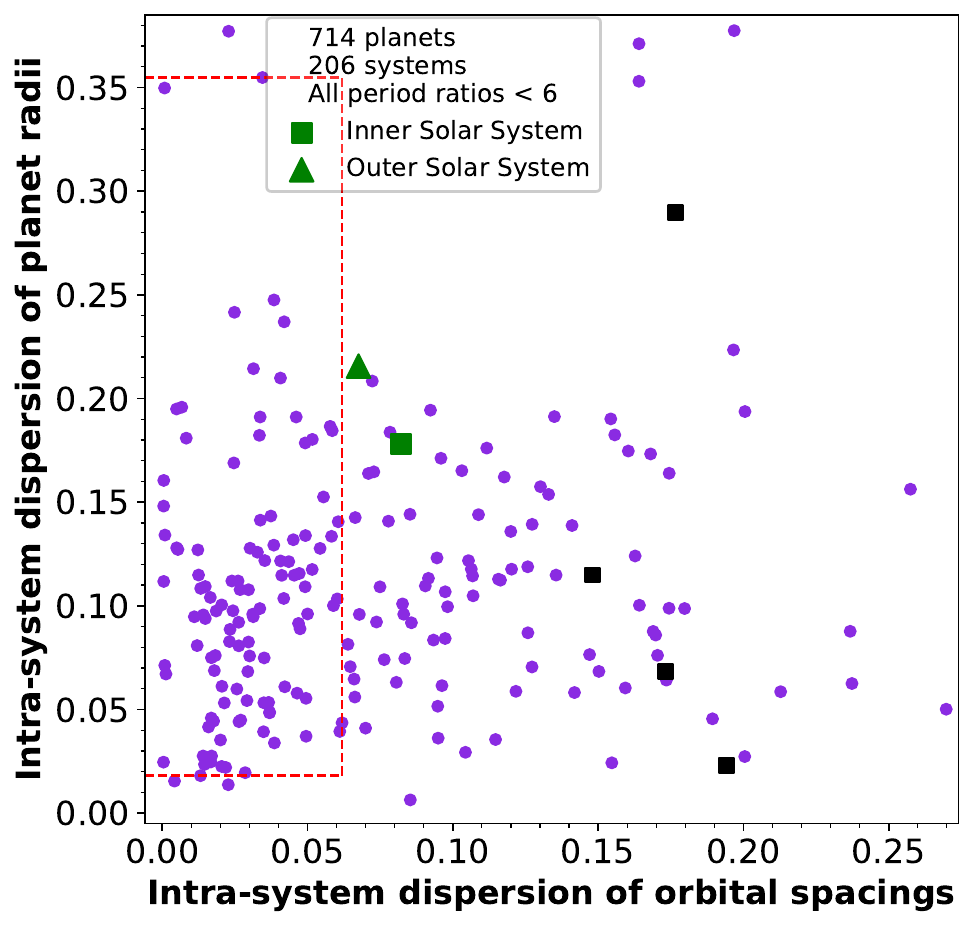}}
\caption{Size dispersion $\sigma_R$ versus spacing dispersion $\sigma_\mathrm{PR}$ (Eq.~\ref{eq:sigma}) of all pairs of adjacent planets in a system. This shows the 206 systems from the radius catalogue in which all adjacent planet pairs have \mbox{PRs < 6}. There is no correlation between the two dispersions. Among the points enclosed by the red lines, there are 51 systems in which all \mbox{PRs < 2}.
The black squares represent the inner regions of five systems that contain a minimum of three inner small planets and at least one outer giant.
Both the inner Solar System (square) with the terrestrial planets and the outer Solar System (triangle) with the four giants possess typical dispersions (Sect.~\ref{sect:size}).} 
\label{fig:std_r_pratio}
\end{figure}

\section{Relations between planet size and spacing} \label{sect:size}

In this section we first examine whether the intra-system similarity of orbital spacings is connected to that of the planetary radii. Then we explore whether the previously reported correlation between period ratios and planetary radii emerges in our sample.

The dispersion of planetary sizes in a system indicates how similar the planets are in terms of their radii $R$. We used the metric given in Eq.~\ref{eq:sigma} and computed $\sigma_R$, where the radii are in units of Earth radii $R_\oplus$. 
We remind that this metric measures the radii in log-space, and therefore it actually implements the ratios and not the differences between the planets' radii and the mean value in a system.     

\begin{figure*}
\sidecaption
\includegraphics[width=12cm]{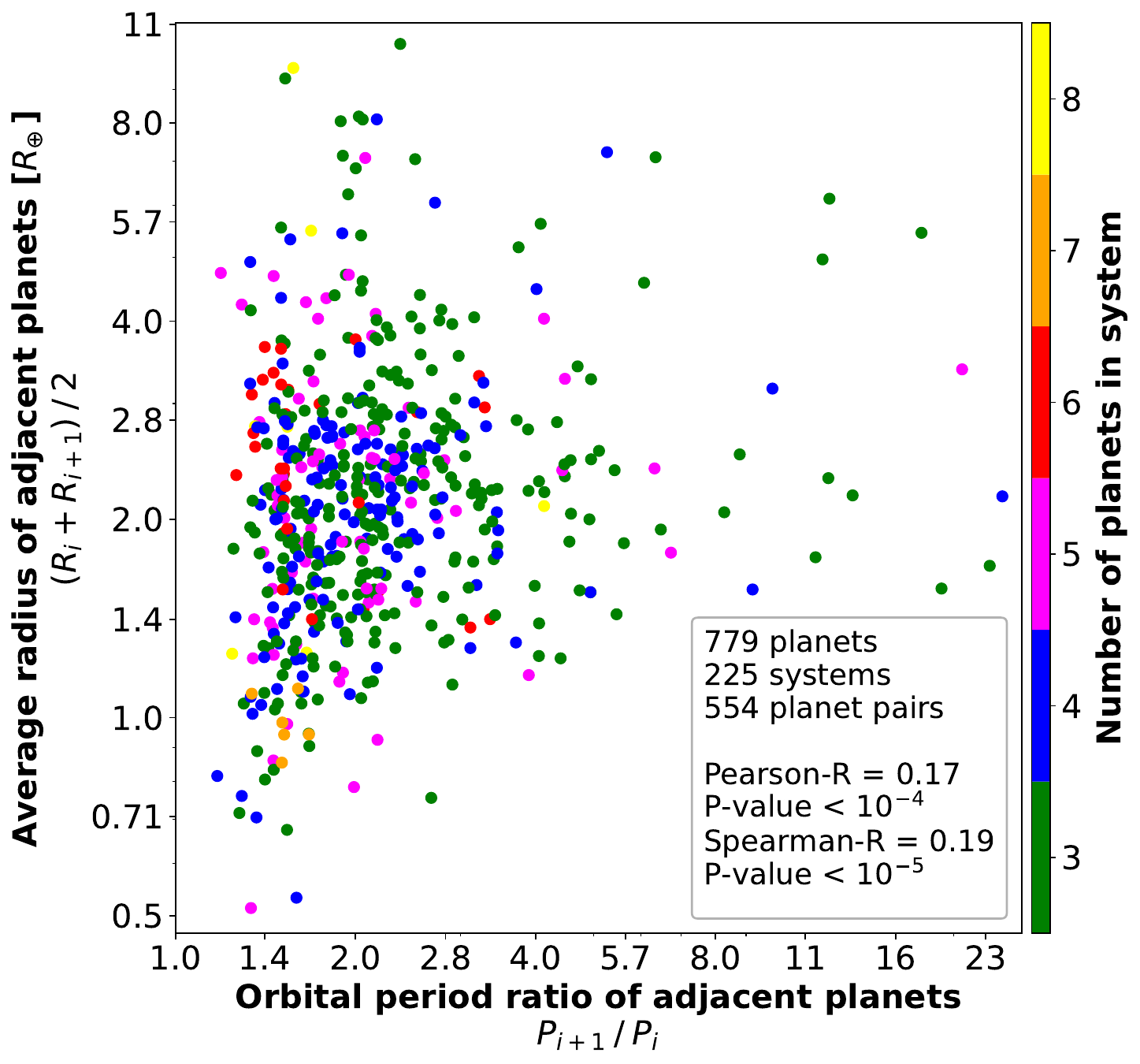}
\caption{Average radii versus orbital period ratios of all 554 pairs of adjacent planets in the radius catalogue. Each point is colour-coded based on the number of planets in the system. There is a weak positive correlation that breaks down when planet pairs with average radii $< 1\, R_\oplus$ are excluded. Very small and very large planet pairs have predominantly small and moderate PRs, respectively, while the intermediate planets span a much wider range of orbital spacings (Sect.~\ref{sect:size}).}
\label{fig:avgr_pratio}
\end{figure*}

We investigated whether there exists a correlation between the intra-system dispersion of orbital spacings and that of planetary sizes. When including all 225 systems from the radius catalogue, the Pearson and Spearman tests indicate a weak positive correlation between the two dispersions $\sigma_R$ and $\sigma_\mathrm{PR}$. Next, we performed the similar tests on different sub-samples and examined whether the relation between the two dispersions changes. 

Our analyses indicate that the correlation diminishes significantly with a decreasing upper PR limit; that is, when we selected the systems in which all pairs of adjacent planets have PRs smaller than the specified limit.
Surprisingly, the correlation between the two dispersions vanishes completely already at an upper PR limit of 6. Figure \ref{fig:std_r_pratio} displays the size dispersions vs. the spacing dispersions in the 206 systems (out of 225) from the radius catalogue in which all the pairs of adjacent planets have $\mathrm{PRs} < 6$. This may suggest that for nearly all the transiting-planet systems, the intra-system dispersion of the planetary sizes is uncorrelated with that of the orbital spacings. 

Figure \ref{fig:std_r_pratio} also shows the corresponding values for the inner and outer Solar System. They resemble the other systems in the sample and have typical spacing dispersions of 0.08 and 0.07, respectively, but larger size dispersions of 0.18 and 0.22. On the contrary, the entire Solar System cannot be encompassed in this figure due to its large size dispersion of 0.60 even though its spacing dispersion is only 0.18. 

In order to perform a more adequate analysis for the inner Solar System, we compared it only to other inner planetary systems. We selected the systems in the period catalogue that contain minimum three inner transiting planets with $R < 9\, R_\oplus$ (i.e. smaller than giant planets) and at least one outer massive planet with $M \sin{i} \geq 50\, M_\oplus$ (following the threshold in \citealp{he23}). There are five such systems of which Kepler-139 and Kepler-65 each have three inner planets and one outer giant. The other three systems are: Kepler-48 with three inner planets and two outer giants, Kepler-90 with seven inner and one outer planet, and HD 191939 with three inner planets and two outer giants. The size and spacing dispersions of the inner planets in each of these systems are marked in Fig.~\ref{fig:std_r_pratio} with black squares, except for Kepler-139 since its inner planets exhibit a very large spacing dispersion of 0.35, thus exceeding the domain of the graph. All five inner systems have larger spacing dispersions than the inner Solar System, but except for Kepler-90, they all have smaller size dispersions. This is further discussed in Sect.~\ref{subsect:undetected}.

As seen in Fig.~\ref{fig:std_r_pratio}, many of the systems in the radius catalogue exhibit an intra-system similarity in orbital spacings and/or planet sizes. 
K2-183 has the greatest spacing dispersion of $\sigma_\mathrm{PR} = 0.58$, which could easily be decreased if an additional planet would be discovered between planet b and c, since these two planets have a high period ratio of 23. Kepler-1311 has the largest size dispersion of $\sigma_R = 0.47$ as well as a high spacing dispersion of $\sigma_\mathrm{PR} = 0.32$. The high period ratio of 18 between planet b and d might indicate that there is an undetected planet between them.

In order to decrease the probability of undetected intermediate planets, we next examined the systems in which all pairs of adjacent planets have $\mathrm{PRs} < 2$. By choosing this low PR limit we assume that there are no undetected planets between the inner- and outermost planet in each of these systems. This selection resulted in 51 systems with a total of 179 planets, which are bordered by the red lines in Fig.~\ref{fig:std_r_pratio}. Evidently, the two dispersions in this sub-sample are not correlated, and all these systems possess small spacing dispersions but span a wide range of size dispersions. This suggests that planets in the same system can have similar orbital spacings even if they are not similarly sized.

As a supplement to the dispersion of sizes and spacings within each system, the relationship between planetary radii and orbital spacings may also be explored on a population level for all pairs of adjacent planets in the sample. \citet{weiss18a} identified a weak positive correlation between the average radii and the orbital period ratios of adjacent planets, which we henceforth refer to as the size-spacing correlation. Their sample comprised all 504 adjacent planet pairs in the CKS multis, of which approximately half are in two-planet systems. This relationship is likely to have an astrophysical origin since it also exists in an underlying unbiased synthetic population \citep{mishra21, weiss23}. Moreover, its strength is shown to diminish due to the biases and limitations of the transit method as well as the completeness of the \textit{Kepler} survey \citep{mishra21}.

We investigated whether this size-spacing correlation emerges in our radius catalogue as well, since it comprises more planet pairs (554) than the CKS multis and does not contain neither unconfirmed planets nor systems with only two planets. The resulting relationship is seen in Fig.~\ref{fig:avgr_pratio} where each point represents a pair of adjacent planets and is colour-coded by the number of planets in the system. There is a very weak positive correlation with Pearson-R = 0.17 (\textit{p}-value $< 10^{-4}$). This coefficient remained unchanged even when we excluded the planet pairs with $\mathrm{PRs} > 4$ in order to decrease the probability of undetected intermediate planets. 

The most prominent features in Fig.~\ref{fig:avgr_pratio} are that small planet pairs with average radii $R < 1.1\, R_\oplus$ have predominantly small period ratios of $\mathrm{PRs} < 1.7$. Conversely, large planets with $R > 4\, R_\oplus$ tend to possess larger PRs, while intermediate-sized planet pairs span a wide range of orbital spacings and reveal no apparent pattern. 

As noted by \citet{weiss18a}, there is a noticeable absence of planet pairs with average radii $< 1\, R_\oplus$ at large orbital period ratios. When we selected from the radius catalogue the 21 planet pairs in this size range, we found a typical value of $\mathrm{PR \leq 2}$, which is in agreement with \citet{weiss18a}. This implies that the vast majority of small-sized planets are in tightly packed configurations. The latter authors also determined from bootstrap models that planets smaller than $1\, R_\oplus$ are in fact detectable at period ratios up to four. Their conclusion was that the lack of such planets with period ratios of $2 < \mathrm{PRs} < 4$ likely has an astrophysical origin. They identified a stronger correlation when examining only the planet pairs with average radii $< 1\, R_\oplus$. In contrast to this result, we found no correlation between the PRs and average sizes of the 21 planet pairs with average $R < 1\, R_\oplus$ in our radius catalogue.

We also tested how excluding these 21 planet pairs affects the planet size-spacing trend, first with no upper PR limit and then with a limit of 4. The conclusions from both these approaches are consistent with one another, revealing that there is no correlation between the average radii and the period ratios in this sub-sample. This implies that the correlation between the sizes and PRs of adjacent planets shown in Fig.~\ref{fig:avgr_pratio} disappears when planet pairs with average \mbox{$R < 1\, R_\oplus$} are excluded.
Only when we additionally excluded planet pairs larger than $4\, R_\oplus$ did a significant, but very weak, correlation emerge, which then increased with decreasing upper limit on the radius. These results may have two implications: Firstly, for planets with radii of $1\,R_\oplus \leq R \leq 4\, R_\oplus$, the orbital period ratios tend to be larger as the planets' radii increase. Secondly, planets with $R > 4\, R_\oplus$ show no preference for their orbital spacings based on the planet size.

\section{Relations between planet mass and spacing} \label{sect:mass}

\begin{figure}
\resizebox{\hsize}{!}{\includegraphics[width=17cm]{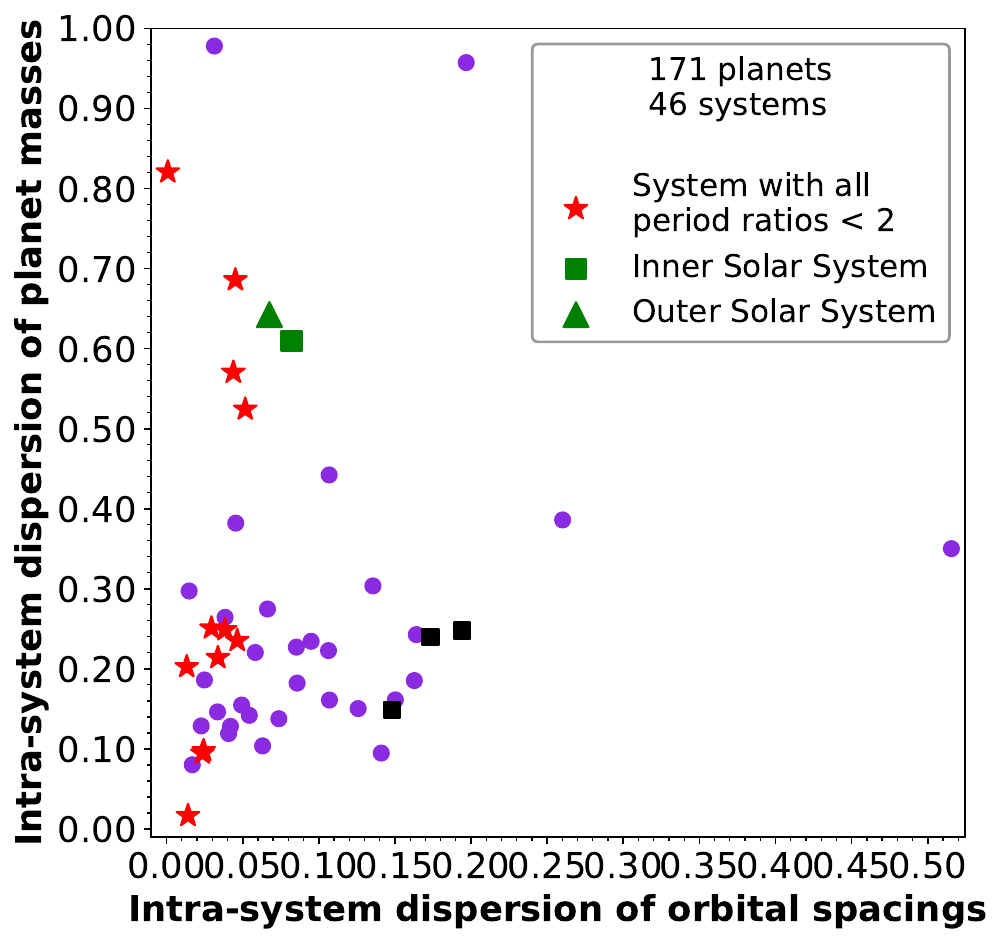}}
\caption{Mass dispersions $\sigma_M$ versus spacing dispersions $\sigma_\mathrm{PR}$ (Eq.~\ref{eq:sigma}) for all 46 systems in the mass catalogue. There is no correlation between the two dispersions. The 12 systems in which all $\mathrm{PRs} < 2$ (star symbols) have small spacing dispersions but span a wide range of mass dispersions. Both the inner and outer Solar System have large mass dispersions.
The black squares correspond to the inner regions of three systems that contain at least three inner small planets and a minimum of one outer giant (Sect.~\ref{sect:mass}).}
\label{fig:std_m_pratio}
\end{figure}

The observed relationships between planetary sizes and orbital spacings likely originate from underlying relationships between the masses and spacings of planets. To explore this, we used the sample in the mass catalogue and conducted the same analyses as in the previous section focusing on the planetary masses instead. This has not been pursued in previous studies on observed data although \citet{mishra21} examined the masses vs. the period ratios of a synthetic planet population.

We focused exclusively on the mass catalogue since we wanted to utilise the planetary masses $M$ and not $M\sin{(i)}$ measurements or values obtained through mass-radius relationships. First, we investigated the intra-system  similarities using Eq.~\ref{eq:sigma}, and the mass dispersions $\sigma_M$ vs. the spacing dispersions $\sigma_\mathrm{PR}$ of all 46 systems are plotted in Fig.~\ref{fig:std_m_pratio}.

The inner and outer Solar System have $\sigma_M = 0.61$ and $\sigma_M = 0.64$, respectively, and are shown in the figure as well. Only four systems have mass dispersions larger than these two values. Meanwhile, the entire Solar System has a very large value of $\sigma_M = 1.73$ and cannot be encompassed in Fig.~\ref{fig:std_m_pratio}. 

Similar to our procedure in Sect.~\ref{sect:size}, we selected the five systems in the period catalogue that contain at least three inner transiting planets and a minimum of one outer giant. Only three out of these systems have true mass values for all their inner planets: HD 191939, Kepler-48, and Kepler-65. The mass and spacing dispersions of the inner regions of these systems are marked in Fig.~\ref{fig:std_m_pratio}, showing that they have larger spacing dispersions but smaller mass dispersions compared to the inner Solar System.

Although the number of systems is not statistically large, our analysis indicates that the two dispersions are not correlated. This result agrees with our conclusions in Sect.~\ref{sect:size} and solidifies our finding that planets in the same system can be similarly spaced even if they do not have similar sizes or masses. To test this idea further, we proceeded as in Sect.~\ref{sect:size} and examined the systems in which all pairs of adjacent planets possess $\mathrm{PRs} < 2$. Thus, we assume that these systems have tightly packed planets without any undetected intermediate planets. There are only 12 such systems (marked in Fig.~\ref{fig:std_m_pratio}) compared to the 51 systems in the radius catalogue. In contrast to their small spacing dispersions, they clearly display much larger mass dispersions, particularly the four systems with $\sigma_M > 0.26$. This is in agreement with the trend in planet size in Fig.~\ref{fig:std_r_pratio} but is more pronounced for the planet mass.

\begin{figure}
\resizebox{\hsize}{!}{\includegraphics[width=17cm]{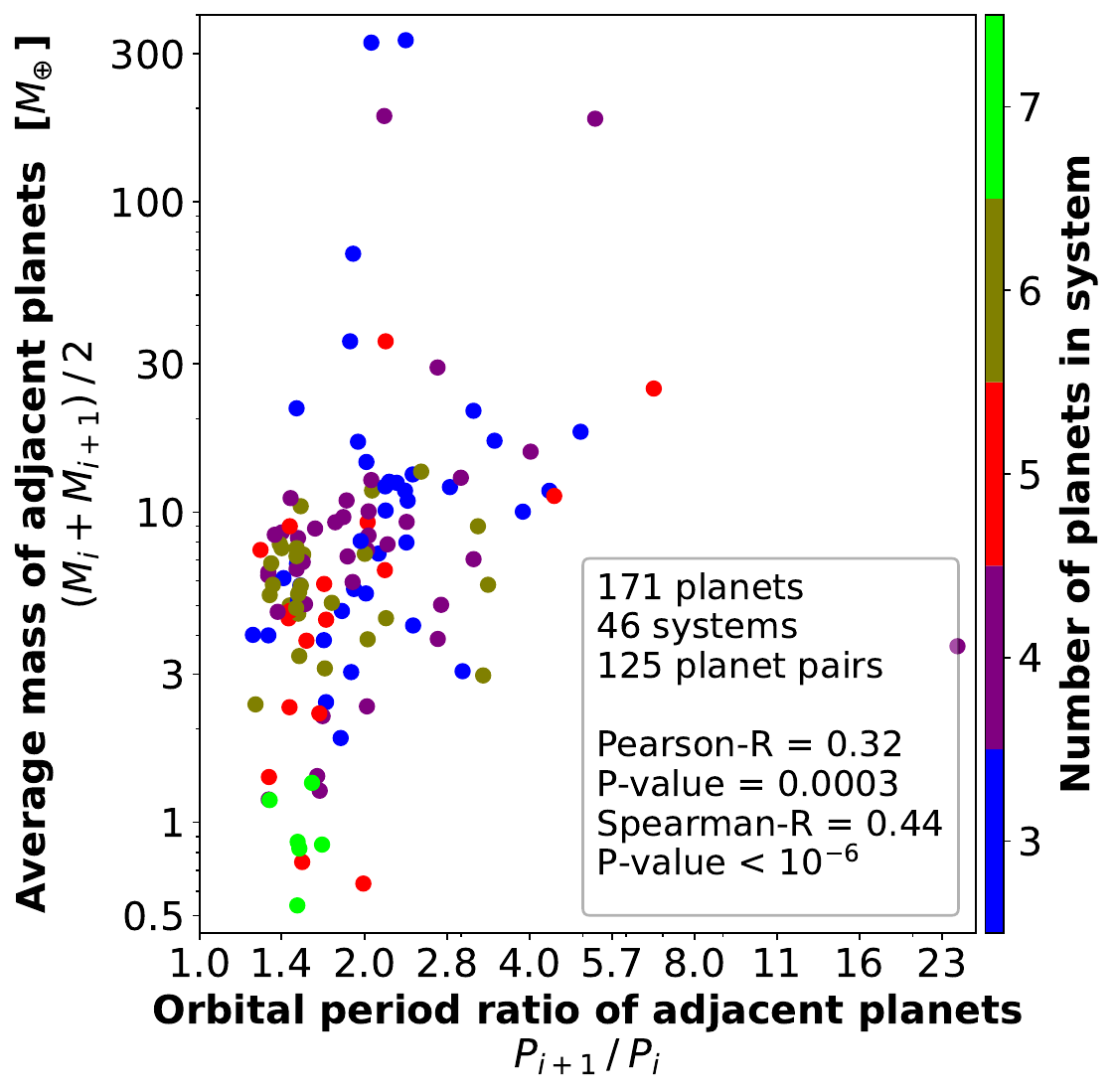}}
\caption{Average mass versus orbital period ratio of every two adjacent planets in the mass catalogue. There is a weak Pearson and a moderate Spearman correlation, which are much higher than in Fig.~\ref{fig:avgr_pratio}. Each point is colour-coded based on the planet multiplicity in the system. The majority of planet pairs in systems with $\geq 5$ planets have small PRs and average masses of $M < 10 \, M_\oplus$.}
\label{fig:avgm_pratio}
\end{figure}

\begin{figure*}
\sidecaption
\includegraphics[width=12cm]{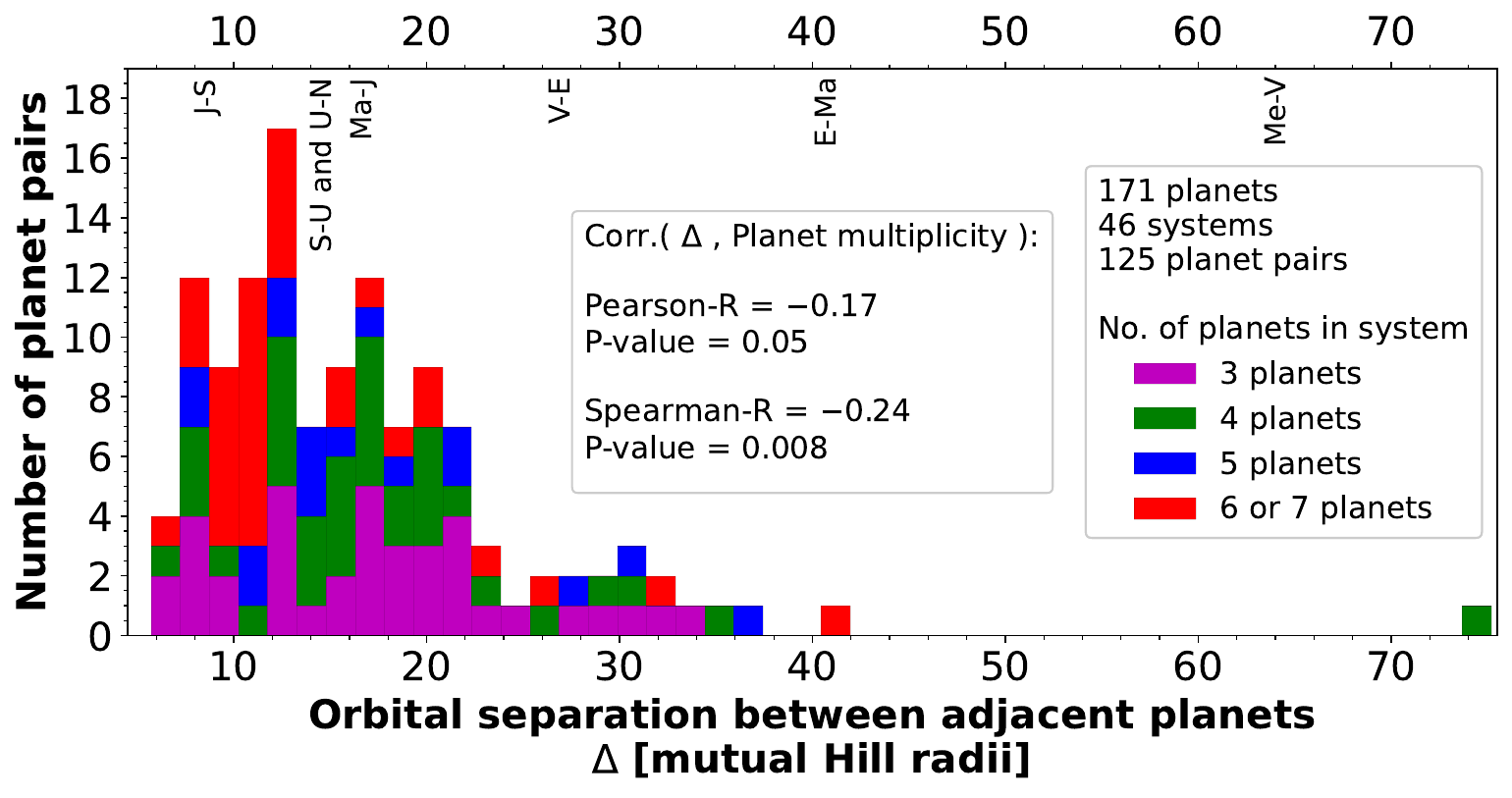}
\caption{Stacked histogram displaying the distributions of orbital separations $\Delta$ (in units of mutual Hill radii) of all 125 pairs of adjacent planets in the mass catalogue shown for each planet multiplicity: systems with three (magenta), four (green), five (blue), and six or seven (red) detected planets. The separations tend to be smaller in high-multiplicity systems.
The planets TOI-561 b and c have the largest mutual separation of $\Delta = 75$. The uppercase letters at the top mark the values of all the adjacent planet pairs in the Solar System (Sect.~\ref{sect:hill}).}
\label{fig:hist_hsep}
\end{figure*}

Kepler-60 is the system with the smallest dispersions in the lower left corner of Fig.~\ref{fig:std_m_pratio}, exhibiting a peas-in-a-pod architecture for both mass and spacing. The systems with the highest and next-highest spacing dispersions are TOI-561 and Kepler-62, respectively, although both the radius and period catalogues contain systems with larger dispersions.
Systems with large $\sigma_\mathrm{PR}$ might harbour undetected intermediate planets, especially if they display small $\sigma_M$ values.

Complementary to focusing on each system individually, we investigated all 125 pairs of adjacent planets in the mass catalogue. The average mass as a function of the orbital period ratio of each pair is displayed in Fig.~\ref{fig:avgm_pratio}. Massive planets, in particular with average $M > 10\, M_\oplus$, tend to have larger PRs than low-mass planets. 
The orbital spacings are more strongly correlated with the planetary masses than with the radii (Fig.~\ref{fig:avgr_pratio}), and both the Pearson and Spearman coefficients are higher. 
Complementarily, in a synthetic population of multi-planetary systems, \citet{mishra21} identified a stronger linear correlation of the PRs with the masses than with the radii of planets. For the first time we investigated and corroborated this result in a sample of detected planets with true mass values. We note, however, that the data sample in Fig.~\ref{fig:avgr_pratio} is almost 4.5 times larger compared to the one presented in Fig.~\ref{fig:avgm_pratio}, although the correlations are significant in both samples.

In the mass catalogue, the planet pair \mbox{TOI-561 b - c} has the highest period ratio of $\approx24$, since planet b is an ultra-short-period planet, while planet c has an orbital period of $\approx11$ days. 
As seen in Fig.~\ref{fig:avgm_pratio}, the majority of planet pairs in systems with $\geq 5$ planets have average masses of $M < 10\, M_\oplus$ and small orbital spacings irrespective of their masses. Both the PRs and masses are small for all the planet pairs in the TRAPPIST-1 system. Two of the four planet pairs with masses $M > 100\, M_\oplus$ orbit Kepler-30, while the other two reside in the WASP-47 system.

\section{Mutual Hill radii separations} \label{sect:hill}

\begin{figure}
\resizebox{\hsize}{!}{\includegraphics[width=17cm]{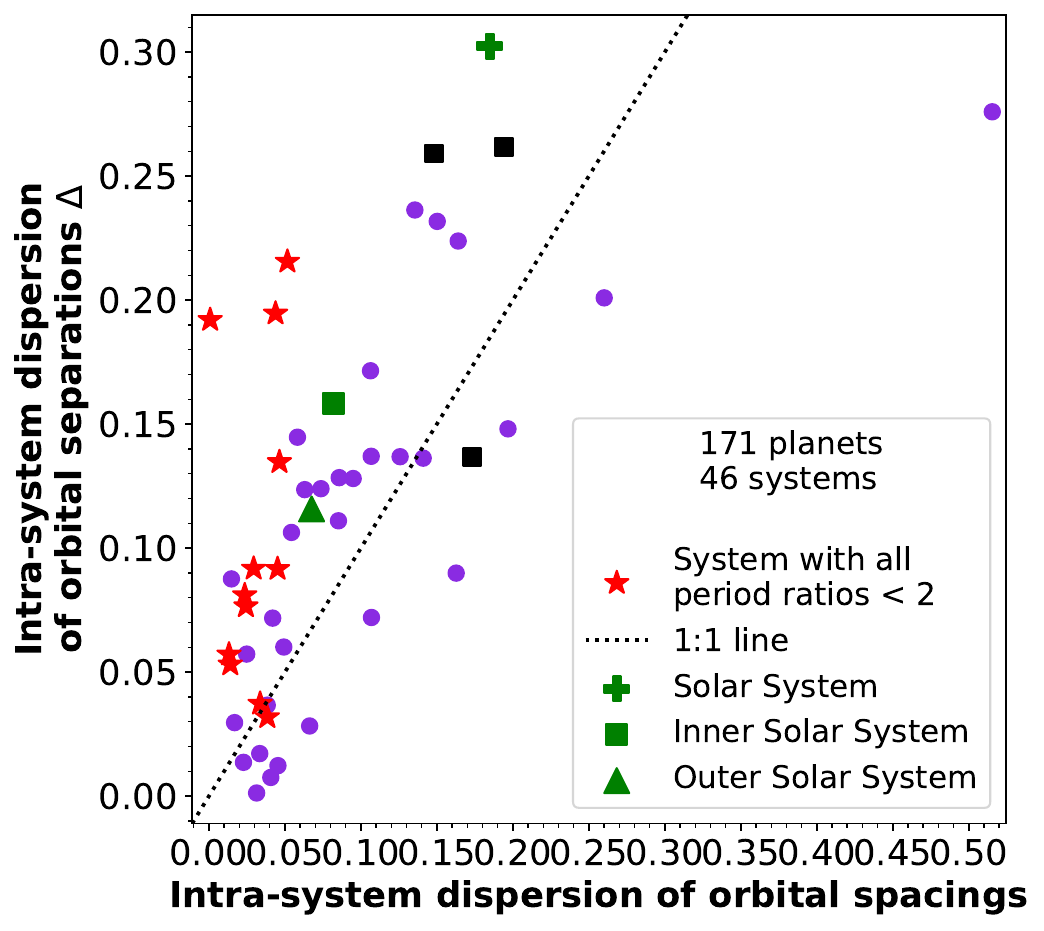}}
\caption{Dispersion of orbital separations $\Delta$ versus the dispersion of PRs between adjacent planets in each of the 46 systems in the mass catalogue. The majority of systems display greater dispersions of separations $\Delta$ than of PRs, and the upper right outlier is the TOI-561 system. The 12 systems (star symbols) in which all $\mathrm{PRs} < 2$ have small PR dispersions but span a wide range of $\Delta$ dispersions. Both the inner and outer Solar System have typical dispersions, while the entire Solar System has greater values. 
The black squares represent the inner regions of three systems that contain a minimum of three inner small planets and at least one outer giant.}
\label{fig:std_hsep_pratio}
\end{figure}

Planets interact gravitationally with each other, implying that larger and more massive planets necessitate wider mutual separations compared to their smaller counterparts. Each planet has its own radius of gravitational influence, denoted as its Hill radius, where its influence dominates over that of the host star. For two adjacent planets with masses $M_i$ and $M_{i+1}$ this can be represented by their mutual Hill radius defined as
\begin{equation*}
    R_H = \left(\frac{M_i + M_{i+1}}{3 M_\star}\right)^{1/3} \frac{a_i + a_{i+1}}{2} \; ,
\end{equation*}
where $a_i$ and $a_{i+1}$ are the semi-major axes of the planets' orbits around their host star with mass $M_\star$.
The orbital separation $\Delta$ between two adjacent planets is given in units of their mutual Hill radius:
\begin{equation} \label{eq:hsep}
    \Delta = \frac{a_{i+1} - a_i}{R_H} \ .
\end{equation}

Is the fundamental orbital spacing between two adjacent planets best represented by their period ratio $P_{i+1}/P_i$ or their separation $\Delta$?
We concluded in Sect.~\ref{subsect:similarity} that adjacent planets in the same system often have similar orbital period ratios, as revealed in Fig.~\ref{fig:pratio25}. We ask now whether this trend might in fact emerge from an underlying similarity of these planets' separations $\Delta$  \citep[cf.][]{weiss18a}. As defined in Eq.~\ref{eq:hsep}, the mutual Hill radii separations depend partly on the semi-major axes and are, therefore, related but not directly proportional to the period ratios. For instance, even if two pairs of adjacent planets have equal period ratios, the pair with the less massive planets has a larger separation $\Delta$. 

In this section we investigate whether the orbital spacings between adjacent planets are more similar in terms of their period ratios or Hill radii separations. Figure~\ref{fig:hist_hsep} shows the distribution of $\Delta$ of all the 125 pairs of adjacent planets in the mass catalogue. Apart from the pair \mbox{TOI-561 b - c} with $\Delta = 75$, the smallest and largest separations are $\Delta = 5.7$ and 41, respectively. 
The values for all seven pairs of adjacent planets in the Solar System are marked in Fig.~\ref{fig:hist_hsep}. All three planet pairs in the inner Solar System have larger separations than the four remaining pairs. Although Mars and Jupiter have the largest PR of 6.3, Mercury and Venus have the largest separation of $\Delta = 63$.

As seen in Fig.~\ref{fig:hist_hsep}, there is a negative correlation between the planet multiplicity and the separations $\Delta$. This corroborates previous findings showing that planets in high-multiplicity systems are more dynamically packed than in lower-multiplicity systems \citep[e.g][]{fabrycky14, pu, weiss18a}.

The total distribution has a median of $\Delta \approx 15$ and a mean of $\Delta \approx 17$, while peaking around $\Delta \approx 12$. These values are in agreement with the results of \citet{pu}, who evaluated the separations $\Delta$ in \textit{Kepler} systems with four, five, and six candidates. 

We investigated whether the spacing similarity trend is more pronounced in units of orbital period ratios (Fig.~\ref{fig:pratio25}) or mutual Hill radii by comparing the standard deviations $\sigma_\mathrm{PR}$ to $\sigma_\Delta$ for all 46 systems in the mass catalogue. These are shown in Fig.~\ref{fig:std_hsep_pratio} along with the values for the inner, outer, and the entire Solar System. 
We find that most of the systems lie above the one-to-one line in the figure, implying that they have planets that are more dispersed in $\Delta$ than in period ratios. This suggests that the orbital spacings between adjacent planets are more similar in units of period ratios than of mutual Hill radii. 

Figure \ref{fig:std_hsep_pratio} also displays the values for the inner transiting planets in each of the three systems: HD 191939, Kepler-48, and Kepler-65, as in Fig.~\ref{fig:std_m_pratio}. Although the inner Solar System possesses a smaller spacing dispersion compared to these systems, its dispersion of $\Delta$ is slightly larger than that of Kepler-48.

Finally, as in Sect.~\ref{sect:mass} we examined the 12 systems in which all the pairs of adjacent planets have $\mathrm{PR} < 2$, as marked with star symbols in Fig.~\ref{fig:std_hsep_pratio}. These systems harbour planets that we consider to be so tightly packed that no intermediate planets can have stable orbits. Although these systems have low period ratio dispersions $\sigma_\mathrm{PR}$, they span a wide range of separation dispersions $\sigma_\Delta$. This supports our conclusion that the intra-system spacing similarity is more pronounced in units of orbital period ratios than of mutual Hill radii separations.

\section{Discussion} \label{sect:discussion}
In this section we consider the impact of observational biases in our data and discuss some of the results and limitations of our analyses.

\subsection{Detection biases and limitations} \label{subsect:bias}
Throughout our work we have considered how the limitations and biases of the transit and RV surveys might affect our analyses. Since 794 out of all the 991 planets in the period catalogue have been discovered using the transit method, the geometrical limitations and detection biases of this technique could play a key role. 
If unaccounted for, observational biases might distort the following five underlying properties: the similarity of adjacent spacings, the intra-system spacing dispersion, and the relationships between the spacings and the planets' sizes, masses, and multiplicities.  

Observational biases might cause planets in a system to remain undetected. This could for instance occur in transit surveys if these planets have high mutual inclinations and are not seen transiting their star. 
\citet{mishra21} showed that both the spacing similarity and the size-spacing trends are diminished by the biases and limitations of the transit method as well as the completeness of the \textit{Kepler} survey.

For example, we can consider a hypothetical system of five planets with orbital periods of 1, 4, 16, 64, and 256 days. The inner three planets are equally sized, while the outer two planets are much larger. The period ratio between all pairs of adjacent planets is four and is, thus, uncorrelated with the planet size. We further consider that \textit{Kepler} has observed this system, but that the planet at 64~days is not transiting. Hence, what we actually detect is a four-planet system in which the two inner pairs of adjacent planets have similar sizes and period ratios of four, whereas the outer planet pair has a larger average size and a period ratio of 16. This observational bias has five erroneous effects: 
\begin{enumerate}[label=\alph*), topsep=2.7pt, itemsep=0.25em]
    \item It renders our observation sample incomplete.
    \item It strengthens the negative correlation between planet multiplicity and period ratios (Sect. \ref{subsect:diversity}).
    \item It diminishes the similarities of adjacent spacings (Sect. \ref{subsect:similarity}).
    \item It increases the intra-system spacing dispersion (Sect. \ref{subsect:intra-system similarity}).
    \item It increases the positive correlation between planetary sizes and period ratios (the size-spacing trend; Sect.~\ref{sect:size}).
\end{enumerate}

In other cases instead of being strengthened, the size-spacing correlation might on the contrary be weakened by detection biases. This could occur if small-sized tightly packed planets remain undetected due to their very shallow transit depths while larger and more widely spaced planets are observed.

As seen in Fig.~\ref{fig:pratio25}, the similarity of adjacent spacings strongly decreases at high orbital period ratios. This feature is likely strengthened by missing intermediate planets. Furthermore, the majority of planets with high PRs have long orbital periods, and the probability of detecting these planets transiting is low. We are, therefore, biased against detecting planet pairs with similar spacings if their PRs are high. This limitation also makes our dataset incomplete and strengthens the negative correlation between planet multiplicity and orbital spacings. 

Taking the above considerations into account, we analysed several data samples from our catalogues specifying different lower and upper PR limits. In order to mitigate the detection biases, in particular the effects of undetected intermediate or outer planets, we decreased the upper PR limit incrementally from 2662 down to 2 in our analyses in \mbox{Sects.~\ref{sect:spacings} - \ref{sect:hill}}.
For instance, in Sect.~\ref{subsect:similarity} we examined how the large PRs in the period catalogue influence the similarity trend of adjacent spacings. In contrast to \citet{jiang20}, we concluded that this trend extends to high $\mathrm{PRs} > 4$ and even up to $\mathrm{PR} = 2662$ as long as all PRs down to 1.7 or less are included. 

As another example, we investigated in Sect.~\ref{sect:size} how the high PRs impact the correlation between the dispersions of planetary sizes and spacings. 
For the 206 systems from the radius catalogue in which all the pairs of adjacent planets have $\mathrm{PRs} < 6$, there is no correlation between the intra-system dispersions of the planetary radii and those of the period ratios. However, including the remaining 19 systems in which one pair of adjacent planets possesses $\mathrm{PR} > 6$, erroneously gives rise to a positive correlation. These high PRs might be caused by detection biases, and therefore we do not consider this correlation to have an astrophysical origin. 

\subsection{System-level analyses}
Previous studies have often used correlation tests in order to investigate common architecture trends across many systems, as for instance in Figs.~\ref{fig:pratio4} and \ref{fig:avgr_pratio} \citep[e.g.][]{weiss18a, jiang20, mishra21, weiss23}. In such analyses either the planets or the planet pairs from all the systems in a sample are tested collectively. This is very efficient on large datasets and can reveal common patterns on a population level. However, this approach is no longer suitable if we want to compare the architecture of one system to that of another, since it does not enable us to keep track of each individual system. 
This is noticeable for instance in Fig.~\ref{fig:pratio25} showing the correlation between every two adjacent spacings for a total of 685 planet pairs. Owing to the large number of data points, the correlation is statistically significant and is ideal for examining the similarities of all the adjacent spacings in the sample.
Nevertheless, for the systems with more than three planets, it is not evident in Fig.~\ref{fig:pratio25} which of the points represent the same system. Therefore, we cannot fully characterise the spacing similarity within an individual system using this correlation test alone.

Hence, a few other approaches have been adopted in system-level studies \citep[e.g.][]{gilbert, goyal, weiss23, mishra23a}. These have applied different metrics to measure specific properties within a planetary system or to characterise its architecture. In this work we used Eq.~\ref{eq:sigma} to quantify the intra-system dispersion of each of the following quantities: the planet radii, the planet masses, the period ratios of adjacent planets, and their Hill radii separations. 
Naturally, each of these various approaches provides its own advantages and limitations that should be considered.

One common weakness for the metrics found in the literature is that they do not take into account the type of system that is being examined, for instance with respect to the planet masses or sizes in the system. In particular, low- and high-mass systems probably correspond to two different classes of systems, as indicated by previous research \citep[e.g.][]{morrison, emsenhuber, jiang23, wang}. Therefore, it might not always be correct to apply the same metric on two different types of systems; for example on a system with low-mass and tightly packed planets as well as on a system that has both inner low-mass planets and outer giants. All the metrics utilised in past studies would indicate a higher degree of intra-system similarity for the low-mass system, especially with respect to planet size and mass. 
For instance, in a forming low-mass system with pairs of adjacent planets with masses $< 40\, M_\oplus$, theoretical research has shown that energy optimisation leads to nearly equal-mass planets in circular and coplanar orbits, when subjected to conservation of angular momentum, constant total mass, and fixed orbital spacings \citep{adams20, adams19}. Hence, in a low-mass system a peas-in-a-pod architecture is energetically favoured over an architecture with high mass dispersion. On the other hand, in a high-mass system with a planet pair of mass $\gtrsim 40\, M_\oplus$, energy optimisation leads to a configuration in which one planet undergoes runaway growth and acquires most of the mass \citep{adams20}. This greatly reduces the intra-system similarity of planet sizes and masses. 

We must, therefore, be cautious when exploring common architecture trends using the same metric on all systems. For this reason, we investigated whether the conclusions presented in Sects.~\ref{sect:size} and \ref{sect:mass} hold true when dividing the systems into different classes and examining each class separately. We selected from the radius catalogue the following four categories of systems: 
\begin{itemize}[topsep=3.1pt, itemsep=0.28em]
    \item Systems in which all planets have $R \leq 2\, R_\oplus$.
    \item Systems that have minimum one planet with $R > 2\, R_\oplus$.
    \item Systems in which all planets have $R \leq 3.5\, R_\oplus$.
    \item Systems with a minimum of one planet with $R > 3.5\, R_\oplus$.
\end{itemize} 
This enabled us to examine the four system classes separately and apply the same metric as in Sect.~\ref{sect:size} in order to measure the intra-system similarities. The results for the four system types are consistent both with each other and with our previous conclusions regarding the correlations between the planetary radii and spacings as well as the relations between their intra-system dispersions. This justifies our initial analyses of the systems in the radius catalogue without dividing them into different classes based on the planets' radii.

We also conducted similar investigations on the systems in the mass catalogue. We divided them into two categories: one comprising the low-mass systems in which all planets have $M < 10\, M_\oplus$ and another one containing all the remaining systems in the catalogue. 
The results from the analyses within each system class are in agreement both with one another and with our assertions in Sect.~\ref{sect:mass}. This supports our initial conclusions drawn without dividing the systems into different classes.

We underline that a system's architecture is dependent not only on its planets' masses and sizes but also on its age and dynamical evolution. 
In a population synthesis experiment, \citet{mishra21} and \citet{weiss23} found that the vast majority of planet pairs possess small period ratios at early times $\lesssim\!3$ Myrs of planet formation. Their PRs increase during the subsequent 100 Myrs due to dynamical \textit{N}-body interactions. The size-spacing correlation and the spacing similarity trend are absent at early times and emerge later over timescales of millions of years due to the dynamical sculpting in the systems \citep{mishra21, weiss23, lammers}. These dynamical encounters lead to planet-planet scattering, merger collisions, or even ejections of planets, thereby increasing the orbital spacings between adjacent planets \citep{mishra21}.
Large, massive planets have likely undergone more collisions, especially mergers, in the past than the small planets in the same system, thereby clearing more space around them. This may explain why they tend to possess larger PRs, while smaller planets can remain more tightly packed \citep{mishra21}.

\subsection{Undetected planets in known systems} \label{subsect:undetected}
In addition to characterising a planetary system, the intra-system orbital spacings and their dispersion may also be used for suggesting where to search for undetected planets in a system, assuming it has a peas-in-a-pod architecture. A high spacing dispersion, especially in a system with planets of similar masses and sizes, might indicate that there is an undetected planet between the two planets with the highest period ratio. 

Focusing first on the systems in the period catalogue that have been discovered by RV surveys, we could identify each pair of adjacent planets that displays a much higher period ratio than the remaining pairs in its system. For example, planets c and d in the HIP 57274 system have an orbital period ratio of 13.5, whereas the ratio between the inner planets b and c is 3.9. Assuming a regular spacing of $\mathrm{PR}\!\approx\!3.7$ between all adjacent planets, a fourth planet may exist between planets c and d at an orbital period of $\approx\!118$~days. Other systems that might contain undetected intermediate planets are, for instance, GJ 163, HD 181433, HD 27894, and HIP 14810. 
We emphasise that a high period ratio is not necessarily indicative of an undetected planet. It could instead be a natural outcome of dynamical interactions and orbital perturbations, such as planet collisions, mergers, scatterings, and ejection of planets \citep{mishra21}.

Furthermore, for the transiting-planet systems, we can supplement the above PR analyses by computing the intra-system dispersions of both the planetary radii and the orbital spacings. Plotting these against each other as in Fig.~\ref{fig:std_r_pratio} provides valuable information about the system architectures and some of the emerging trends.
In particular, a system with a large spacing dispersion and a small size dispersion is likely to harbour an undetected planet between the two planets with the highest period ratio. Some examples are Kepler-1311, K2-183, Kepler-198, Kepler-286, Kepler-311, Kepler-385, TOI-561, Kepler-62, and Kepler-126. The latter is a three-planet system where the orbital period ratio of the inner and outer planet pair is 2.1 and 4.6, respectively. Assuming that all planets in this system are uniformly spaced with $\mathrm{PR}\approx2.1$, a fourth planet might hide between planet c and d at an orbital period of $\approx\!47$~days. 

In Sect.~\ref{sect:size} we examined the five systems in the period catalogue that contain minimum three inner transiting planets and at least one outer giant. Except for Kepler-90, all the outer giants have been discovered by RVs and have larger spacings than the inner planets. Therefore, these systems exhibit larger spacing dispersions than the majority of the systems in our sample, in agreement with the results of \citet{he23} based on the spacing gap complexities in selected \textit{Kepler} systems. We also computed the dispersions of the spacings, radii, and masses of only the inner planets in these five systems, as shown in Figs.~\ref{fig:std_r_pratio} and \ref{fig:std_m_pratio}. Compared to the rest of the systems and the inner Solar System, they have typical values of $\sigma_R$ and $\sigma_M$, but high values of $\sigma_\mathrm{PR}$. These results corroborate \citet{he23} who concluded that inner transiting-planet systems with high spacing gap complexities are good predictors of outer giant planets, whereas inner systems with low gap complexities generally do not have long-period giants. This may indicate that the giant planet has had a major influence on the system during its formation and subsequent evolution \citep{he23}.

In addition to the systems with large spacing dispersions, systems with very small dispersions might contain undetected planets as well. In such systems searches could be made for interior and/or exterior planets spaced uniformly from their neighbours. For instance, the three planets orbiting Kepler-398 display the smallest spacing dispersion in the radius catalogue, and both pairs of adjacent planets have period ratios of 1.67. If there would be an undetected exterior planet with a similar spacing, it would have an orbital period of 19~days. Other systems with small spacing dispersions and hence potentially undetected planets are, for example, YZ Cet, Kepler-207, Kepler-229, and Kepler-249. 
However, these theoretical considerations are probably more complicated in practice. By examining 64 \textit{Kepler} systems with a minimum of four planets, \citet{millholland22} found that a hypothetical outer planet with the same radius and orbital spacing as its inner adjacent neighbour should be detectable, if it exists, in \mbox{35-44 \%} of these systems. 
The lack of detection of additional outer planets in these systems may indicate that there is a truncation in the underlying system architectures at \mbox{$\sim$100–300} days, or that these outer planets have smaller radii, larger orbital spacings, or higher mutual inclinations \citep{millholland22}.
 
\section{Conclusion} \label{sect:conclusion}

In this work we have investigated 282 multi-planetary systems with a total of 991 confirmed planets and 709 pairs of adjacent planets. We have examined the orbital spacings and their relationships with planet size and mass, conducting both intra- and inter-system analyses.
Our main findings are: 
\begin{enumerate}[topsep=0.3em, itemsep=0.2em]
\item The majority of the systems in our sample exhibit an intra-system similarity in orbital spacings and/or planet sizes.
\item In contrast to previous studies, we have identified a similarity of adjacent orbital spacings not only for $\mathrm{PRs} < 4$ but also for $1.17 < \mathrm{PRs} < 2662$ \citep[cf.][]{weiss18a, jiang20, mamonova}.
\item We have found the intra-system spacing similarity to be more pronounced in units of orbital period ratios than of mutual Hill radii separations $\Delta$.
\item Compared to \citet{weiss18a}, we have identified a weaker positive correlation between the PRs and the average radii of adjacent planets.
\item The aforementioned correlation disappears when planet pairs with $R < 1\, R_\oplus$ are excluded. 
\item We have found the observed PRs to be much more strongly correlated with the planets' masses than with their radii. 
\item Our analyses show no correlation between the intra-system dispersions of the orbital spacings and those of the planetary sizes or masses. Planets in the same system can be similarly spaced even if they do not have similar sizes or masses.
\item The inner and outer Solar System have similar values for their intra-system dispersions, both in planetary sizes, masses, orbital spacings, and $\Delta$. These values range from moderate to large compared to the systems in our catalogues.
\item Compared to the inner regions in other five systems with outer giants, the inner Solar System possesses the lowest spacing dispersion.
\end{enumerate}

Supplementing current data with observations from long-duration surveys, such as the PLATO mission \citep{rauer}, will help us obtain more accurate measurements and determine the true planet multiplicity in systems. Future studies will be better equipped to characterise and compare systems and their multi-faceted architectures. This will increase our knowledge of the formation and evolution of planetary systems, including the Solar System.

\begin{acknowledgements}
We thank the anonymous referee for the very helpful and insightful comments and review of our manuscript. 
This research has made use of the NASA Exoplanet Archive, which is operated by the California Institute of Technology, under contract with the National Aeronautics and Space Administration under the Exoplanet Exploration Program.
\end{acknowledgements}
\bibliographystyle{aa}
\bibliography{bibliography} 
\end{document}